\documentclass[twocolumn,amssymb,amsmath,superscriptaddress,nofootinbib,aps,prd,preprintnumbers,floatfix]{revtex4-1} 
\pdfoutput=1

\usepackage[T1]{fontenc}
\usepackage[utf8]{inputenc}
\usepackage{amsmath,amssymb,amsfonts,mathtools,bm}
\usepackage[final]{graphicx}
\usepackage[usenames,dvipsnames]{xcolor}
\usepackage{booktabs}
\usepackage{relsize}
\usepackage{lineno}
\usepackage{comment}
\usepackage{multirow}
\usepackage{siunitx}
\usepackage{enumitem}
\usepackage{hyperref}
\usepackage{rotating}
\usepackage{dsfont}

\usepackage[normalem]{ulem}
\usepackage{bbding} % checkmarks, crosses

\def\Babar{{\mbox{\slshape B\kern-0.1em{\smaller A}\kern-0.1em B\kern-0.1em{\smaller A\kern-0.2em R}}}}

\newcommand{\phaMinus}{\phantom{-}}

\newcommand{\alphas}{\alpha_\text{s}}
\newcommand{\alphae}{\alpha_\text{e}}

\newcommand{\TeV}{\,\text{TeV}}

\newcommand{\eq}[1]{\eqref{eq:#1}}
\newcommand{\fig}[1]{Fig.~\ref{fig:#1}}
\newcommand{\tab}[1]{Tab.~\ref{tab:#1}}
\newcommand{\tr}{\mathrm{tr}}
\newcommand{\hc}{\text{h.c.}}

\newcommand{\Sec}[1]{Sec.~\ref{sec:#1}}
\newcommand{\App}[1]{App.~\ref{sec:#1}}

\newcommand{\MPl}{M_\text{Pl}}
\newcommand{\dt}{{\widetilde{\delta}}}

\newcommand{\BM}[1]{$\mathbf{BM_{#1}}$}

\newcommand{\Cnine}{{C_9^{\mu}}}
\newcommand{\Cten}{{C_{10}^{\mu}}}
\newcommand{\Cnt}{{C_{9,10}^{\mu}}}
\newcommand{\Cninepr}{{C_9^{\prime \mu }}}
\newcommand{\Ctenpr}{{C_{10}^{\prime \mu }}}
\newcommand{\Cntpr}{{C_{9,10}^{\prime \mu }}}
\newcommand{\Cntprnopr}{{C_{9,10}^{(\prime) \mu}}}

\newcolumntype{G}{>{\columncolor{LightGray}}c}

\begin{document}

\title{
\texorpdfstring{$\bm{B}$}{B}-Anomalies from flavorful \texorpdfstring{$\bm{U(1)^\prime}$}{U(1)'} extensions, safely
}

\author{Rigo Bause}
\email{rigo.bause@tu-dortmund.de}
\author{Gudrun Hiller}
\email{ghiller@physik.uni-dortmund.de}
\author{Tim H\"ohne}
\email{tim.hoehne@tu-dortmund.de}
\affiliation{Fakult\"at Physik, TU Dortmund, Otto-Hahn-Str.4, D-44221 Dortmund, Germany}
\author{Daniel~F.~Litim}
\email{d.litim@sussex.ac.uk}
\affiliation{Department of Physics and Astronomy, University of Sussex, Brighton, BN1 9QH, U.K.}
\author{Tom Steudtner}
\email{tom2.steudtner@tu-dortmund.de}
\affiliation{Fakult\"at Physik, TU Dortmund, Otto-Hahn-Str.4, D-44221 Dortmund, Germany}
\affiliation{Department of Physics and Astronomy, University of Sussex, Brighton, BN1 9QH, U.K.}

\preprint{DO-TH 21/18}

\begin{abstract}
$U(1)^\prime$ extensions of the standard model with generation-dependent couplings to quarks and leptons are investigated as an  explanation of anomalies in rare $B$-decays, with an emphasis on stability and predictivity up to the Planck scale.
To these ends, we introduce
three generations of vector-like standard model singlet fermions, 
an enlarged, flavorful scalar sector,  
and, possibly, right-handed neutrinos, 
all suitably charged under the $U(1)^\prime$ gauge interaction.
We identify several gauge-anomaly free 
benchmarks consistent with $B_s$-mixing constraints, with hints for electron-muon universality violation, and the  global $b \to s$ fit.
We further investigate the complete two-loop running of gauge, Yukawa and quartic couplings 
up to the Planck scale  to constrain low-energy parameters and enhance the predictive power.
A characteristic of models is  that the $Z^\prime$ with TeV-ish mass predominantly decays to invisibles, \textit{i.e.}~new fermions 
or neutrinos.
$Z^\prime$-production can be studied at a future muon collider.
While benchmarks feature 
predominantly 
left-handed couplings $C_9^{\mu}$ and $C_{10}^{\mu}$,  right-handed ones can be accommodated as well.  
\end{abstract}

\maketitle
\tableofcontents
\flushbottom

\section{\bf Introduction}

A key feature of  the $SU(3)_C \times SU(2)_L \times U(1)_Y$ Standard Model (SM) is lepton universality, which states that different generations of leptons only differ by their masses but  interact identically otherwise.
In  recent years, however, lepton universality has come under growing pressure \cite{Bifani:2018zmi}.
 Specifically, ratios $R_{H}$   \cite{Hiller:2003js} for $H=K,K^*,\phi,X_s,\cdots$ of branching fractions of flavor changing neutral current (FCNC) transitions  
\begin{equation}\label{eq:RKratiodef}
    R_{H} = \frac{\int_{q_\text{min}^2}^{q_\text{max}^2} \frac{\text{d}\mathcal{B}(B\rightarrow H \mu^+ \mu^-)}{\text{d}q^2} \text{d}q^2}{\int_{q_\text{min}^2}^{q_\text{max}^2} \frac{\text{d}\mathcal{B}(B\rightarrow H e^+ e^-)}{\text{d}q^2} \text{d}q^2} 
\end{equation}
have been measured below unity, 
suggesting that new physics beyond the SM 
treats electrons and muons rather differently.
Specifically, the most recent measurement 
\begin{equation}
R_K\left(1.1 \, \text{GeV}^2 < q^2 < 6.0 \,  \text{GeV}^2 \right)=0.846^{+0.044}_{-0.041}
\end{equation}
by the  LHCb-collaboration~\cite{Aaij:2021vac} 
evidences a conflict with  universality
at $3.1\,\sigma$, where $q^2$ denotes the dilepton invariant mass-squared.  A similar suppression of muon-to-electron rates is measured
in $R_{K^*}$~\cite{Aaij:2017vbb} at about $(2-3) \sigma$, and, with larger uncertainties, in $b$-baryon decays $R_{pK}$~\cite{LHCb:2019efc}.
Together with additional tensions with the SM predictions in branching ratios and angular observables~\cite{Aaij:2020nrf,Aaij:2020ruw} 
in $B\rightarrow K^{(*)} \mu^+ \mu^-$ decays these are  collectively called the $B$-anomalies.

Explanations for the $B$-anomalies can be achieved in extensions of the SM 
with an additional, flavorful $U(1)^\prime$ gauge symmetry 
through tree-level exchange of a heavy (electroweak mass or above)  $Z'$ vector boson, see for instance \cite{Crivellin:2015lwa,Bhatia:2017tgo,King:2018fcg,Falkowski:2018dsl,Ellis:2017nrp,Altmannshofer:2013foa,Allanach:2021kzj,Allanach:2020kss,Bonilla:2017lsq,Bian:2017rpg,Cen:2021iwv,Greljo:2021xmg}.
In these settings, a violation of lepton universality   arises naturally through generation-dependent $U(1)'$ couplings to quarks and leptons.
It is worth noting that  the flavor data requires quite sizable couplings.
In fact, explaining $R_{K^{(*)}}$ with  a tree-level mediator and couplings  of order unity points towards a scale of about $40$ TeV \cite{Hiller:2021pul},
while a  generic lower bound on the heavy $Z^\prime$ mass is roughly $5$ TeV \cite{Sirunyan:2021khd}.
The minimal  $U(1)^\prime$ charge assignments 
are to the left-handed muon $(F_{L_2})$ and to the left-handed $b$-quark $(F_{Q_3})$. 
The FCNC quark vertex also requires flavor mixing via
a CKM factor $V_{tb}^{\phantom{*}} V_{ts}^*$. 
Altogether, we obtain
\begin{equation}\label{eq:g4bound}
g_4^2 (\mu_0=5 \text{TeV}) \sim  
\frac{(5 \text{TeV}/40 \text{TeV})^2}{V_{tb}^{\phantom{*}} V_{ts}^* F_{L_2} F_{Q_3} }\sim \frac{0.4}{F_{L_2} F_{Q_3}}\,,
\end{equation}
where $g_4$ denotes the $U(1)^\prime$ gauge coupling.
Let us  explain why  \eq{g4bound}  entails a loss of predictivity  below Planckian energies.  
It is well-known that the
running of the gauge coupling
${d \alpha_4}/{d \ln \mu } =  B_4 \,\alpha_4^2>0$  with $\alpha_4=g_4^2/(4 \pi)^2$ 
implies a Landau pole at high energies.
The coefficient $B_4$ is bounded from below by the minimal amount of $U(1)'$ charges
required to explain the $B$-anomalies and to avoid gauge anomalies.
Together with \eq{g4bound} this entails  a lower bound on $B_4\cdot \alpha_4(\mu_0)$
and  an upper bound
 on the Landau pole  below the Planck scale,
\begin{equation}\label{eq:Landau}
\mu_{\rm L}\lesssim  10^{10}\, {\rm TeV}\,.
\end{equation}
If more $U(1)^\prime$ charge carriers are present, the scale \eq{Landau} is lowered further.
 We conclude that  generic explanations of the $B$-anomalies with $U(1)'$ models 
  come bundled with the limitation \eq{Landau}, which requires a cure of its own prior to the Planck scale.

In this work we address the $B$-anomalies from the viewpoint of predictivity and stability up to the Planck scale,
and  put forward new  $Z'$ extensions of the SM which simultaneously circumvent  \eq{Landau} and the notorious in- or metastability of the Higgs.
At weak coupling, it is well-known that Landau poles can be delayed or removed
with the help of  Yukawa interactions, as these,  invariably, slow down the growth of gauge couplings 
\cite{Bond:2016dvk,Bond:2018oco}. 
It is therefore  conceivable that  SM extensions may be found where quantum fluctuations delay putative Landau poles towards  higher energies, possibly even beyond the Planck scale where quantized gravity  is expected to wreak havoc. 
A virtue of this setup are new theory constraints on  matter fields beyond the SM, their types of interactions, and the values of their couplings.
Equally important is that instabilities of the quantum vacuum can be avoided  \cite{Hiller:2019mou,Hiller:2020fbu}. 
These new constraints are complementary to phenomenological ones and lead to an increased predictivity.

Previous searches for
Planck-safe SM  extensions have looked into  explanations for the $(g-2)_{\mu,e}$ anomalies    \cite{Hiller:2019mou}, aspects of flavor through Higgs portals and their phenomenology \cite{Hiller:2020fbu,Bissmann:2020lge,PlanckSafeQuark},
and  settings where the Yukawa-induced slowing-down leads to
  outright fixed points  at asymptotically high energies
\cite{Litim:2014uca,Bond:2017wut,Bond:2017lnq,Bond:2017suy,Bond:2017tbw,Buyukbese:2017ehm,Kowalska:2017fzw,Bond:2019npq,Fabbrichesi:2020svm,Bond:2021tgu}.
General conditions  for the latter  have been derived \cite{Bond:2016dvk,Bond:2018oco},  including  concrete requirements for matter fields and their interactions  in simple \cite{Litim:2014uca,Bond:2017tbw,Kowalska:2017fzw,Buyukbese:2017ehm,Bond:2019npq,Fabbrichesi:2020svm,Bond:2021tgu}, semi-simple \cite{Bond:2017wut,Bond:2017lnq,Hiller:2019mou,Hiller:2020fbu} and supersymmetric \cite{Bond:2017suy} gauge theories; see 
\cite{Martin:2000cr,Abel:2013mya,Litim:2015iea,Gies:2016kkk,McDowall:2018ulq,Schuh:2018hig,Alkofer:2020vtb,Gies:2020xuh} for further aspects and directions.
Hence, to achieve the  desired features, flavorful BSM sectors with both fermions and scalars are instrumental.

With these goals in mind,
we introduce an additional $U(1)'$ gauge group, 
three generations of vector-like SM singlet fermions, an enlarged and flavorful scalar sector, and up to three right-handed neutrinos,  
all charged under the $U(1)^\prime$. By and large, our choices are motivated by formal considerations~\cite{Litim:2014uca,Bond:2016dvk,Bond:2018oco}
and concrete earlier models  \cite{Hiller:2019mou,Hiller:2020fbu}. 
The $U(1)'$ interactions will be key to explain the $B$-anomalies, the new Yukawa and quartics play an important role
in delaying putative Landau poles, while the portals are relevant to stabilize the Higgs.

This paper is organized as follows.
In \Sec{Models} we  introduce new families of $Z'$ models and discuss theoretical and experimental constraints. This includes anomaly cancellations, kinetic mixing, and mixing and stability of the scalar potential.
In \Sec{B-Anomalies} we address the $B$-anomalies by confronting $Z^\prime$ model parameters to the model-independent outcome of a global fit to current $b \to s$ data, and identify  benchmark scenarios.
In  \Sec{Planck-Safety}, we map out the parameter space of benchmark models to find settings which evade putative Landau poles and remain  predictive up to the Planck scale.
In  \Sec{Discussion} we investigate phenomenological implications of our models at colliders and beyond. We conclude in \Sec{Conclusion}.
Some technicalities are relegated to appendices which cover the derivation of the Landau pole bound \eq{Landau} (App.\ref{sec:Landau}), 
flavor rotations (App.\ref{sec:chargeMixDetails}), aspects of $B_s$-mixing constraints (App.\ref{sec:AppBSMixing}), and the necessity or otherwise for right-handed neutrinos (App.\ref{sec:AppRHN}).

\section{\bf  \texorpdfstring{$\bm{Z^\prime}$}{Zprime} Model Set-Up}
\label{sec:Models}

We introduce the BSM model framework in Sec.~\ref{sec:fields} and discuss gauge anomaly cancellation conditions in Sec.~\ref{sec:ACC}, as well as
the consistency conditions for the Yukawa sector in Sec.~\ref{sec:YukInv}.
Kinetic mixing between  $U(1)^\prime$ and the hypercharge boson is discussed in Sec.~\ref{sec:KinMix}, and scalar symmetry breaking and mixing in
Sec.~\ref{sec:SSB}.

\subsection{Fields and Interactions \label{sec:fields}}

\begin{table}[h!]
  \centering
  \renewcommand*{\arraystretch}{1.2}
  \setlength\arrayrulewidth{1.3pt}
   \setlength{\tabcolsep}{1pt}
  \begin{tabular}{|lc|ccccc|}
    \hline
    Field & &  Gen. & $\, U(1)_Y\, $ & $\, SU(2)_L \, $ & $\, SU(3)_C\, $ & $\, U(1)'\, $ \\ \hline
    SM fermions & $Q_i$ & 3 & $+\frac16$ & $2$ & $3$ & $F_{Q_i}$ \\ 
    & $L_i$ & $3$ & $-\frac12$ & $2$ & $1$ & $F_{L_i}$ \\ 
    & $U_i$ & $3$ & $+\frac23$ & $1$ & $3$ & $F_{U_i}$\\ 
    & $D_i$ & $3$ & $-\frac13$ & $1$ & $3$ & $F_{D_i}$\\ 
    & $E_i$ & $3$ & $-1$ & $1$ & $1$ & $F_{E_i}$\\ 
    Higgs scalar & $H$ & $1$ & $+\frac12$ & $2$ & $1$ & $F_H$ \\ \hline
    BSM fermions & $\nu$ & $3$ & $0$ & $1$ & $1$ & $F_{\nu_i}$\\
    & $\psi_L$ & $N_f$ & $Y_\psi$ & $r_2$ & $r_3$ & $F_{\psi_L}$\\
    & $\psi_R$ & $N_f$ & $Y_\psi$ & $r_2$ & $r_3$ & $F_{\psi_R}$\\
    BSM scalars & $S$ & $N_f\times N_f$ & $0$ & $1$ & $1$ & $F_S$\\ 
    & $\phi$ & $1$ & $0$ & $1$ & $1$ & $F_\phi$\\ \hline
  \end{tabular}
  \caption{SM and BSM fields with multiplicities (number of generations) and  representations under  $U(1)_Y \times SU(2)_L \times SU(3)_C \times U(1)'$.
  For our models, we also employ 
   \eq{vector-like} and \eq{fix}, as well as \eq{KaonBound}, \eq{ElectronBound}, \eq{QuarkMassYCCs} and \eq{FH=0},  see text for further details. }
  \label{tab:fields}
\end{table}

We consider $Z^\prime$ extensions of the  
SM with generation-dependent
$U(1)'$ charges $F_{X_i}$ for the SM quarks $(X=Q,\,U,\,D)$, leptons $(X=L,\,E)$ as well as three right-handed neutrino fields $(X=\nu)$. 
Moreover, the SM Higgs $H$ and a $N_f \times N_f$ BSM scalar singlet $S$ are included, 
as well as $N_f$ BSM fermions $\psi_{L,R}$ which carry universal $U(1)'$ charges.  
The $U(1)'$ symmetry is broken spontaneously by a BSM scalar $\phi$, generating a heavy mass for the $Z'$ boson. 
This requires $\phi$ to be a singlet under the SM gauge group,  with $F_\phi \neq 0$.
All matter fields and representations under the ${U(1)_Y  \times SU(2)_L \times SU(3)_C \times U(1)'}$ gauge group are given in \tab{fields}.

Furthermore, we assume the BSM fermions $\psi_{L,R}$  to be vector-like and to come in three generations as well as $S$ to be uncharged under the $U(1)^\prime$ gauge group
\begin{equation}\label{eq:vector-like}
    N_f = 3\,, \qquad F_\psi = F_{\psi_L} = F_{\psi_R}\,, \qquad F_{S} = 0 \,.
\end{equation}
We also restrict the analysis to  BSM fermions which are complete SM singlets
\begin{equation} \label{eq:fix}
    Y_\psi = 0, \qquad r_2 = 1, \qquad r_3 = 1.
\end{equation}
For phenomenological reasons we also employ the following charge assignments:
{\it i)}, to avoid severe constraints  from $K_0$-$\bar{K}_0$-oscillations \cite{Zyla:2020zbs} we choose 
\begin{equation}\label{eq:KaonBound}
F_{Q_1} = F_{Q_2}, \qquad F_{D_1} = F_{D_2} \, , 
\end{equation}
such that $s$ and $d$-quarks couple universally and do not induce FCNCs.
{\it ii)}, due to strong experimental constraints from LEP and other electroweak measurements (cf.~\cite{Ellis:2017nrp}),
we forbid couplings of the $Z^\prime$ to electrons, 
\begin{equation}\label{eq:ElectronBound}
    F_{L_1} = 0\,, \qquad F_{E_1} = 0 \,.
\end{equation}
In the following we restrict ourselves to \eq{vector-like}-\eq{ElectronBound}.

The Yukawa sector 
\begin{equation}\label{eq:Yuk}
\begin{aligned}
    - \mathcal{L}_\text{yukawa} & =  Y^d_{ij}\, \overline{Q}_i  H D_j + Y^u_{ij}\, \overline{Q}_i  \widetilde{H} U_j + Y^e_{ij}\, \overline{L}_i  H E_j \\ 
    &\phantom{=}  + Y^\nu_{ij}\, \overline{L}_i  \widetilde{H} \nu_j  +  y\, \overline{\psi}_{L i} \,S_{ij} \, \psi_{Rj} + \text{h.c.}
\end{aligned}
\end{equation}
consists of the SM interactions and the right-handed neutrino coupling $Y^\nu$ to the Higgs, as well as a pure BSM Yukawa vertex. 
The latter is described by a single coupling $y$, which is protected by a $U(3)_{\psi_L} \!\times\! U(3)_{\psi_R}$ BSM flavor symmetry that is only softly broken. 
This requires the $F_{\psi_{L,R}}$ charges to be universal.
Moreover, additional Yukawa interactions between SM or right-handed neutrino fields and the BSM sector of $\psi_{L,R}$ and $S$ are forbidden by this symmetry. 
On the other hand, for models with $F_\phi = \pm 2\,F_{\nu_i}$, Majorana-like Yukawa couplings $\rho_i^{\pm}$ of the right-handed neutrinos
\begin{equation}
    - \mathcal{L}_\text{majorana} = \rho^+_i\, \phi^\dagger  (\nu_i \varepsilon \nu_i) + \rho^-_i\, \phi \,(\nu_i \varepsilon \nu_i) + \hc
\end{equation}
are implied where $\varepsilon$ denotes the Levi-Civita tensor contracting spinor indices.
However, such scenarios will be avoided in this work. 
Overall, while the SM flavor symmetry is explicitly broken by its Yukawa interactions, the BSM sector remains unaffected.
Non-vanishing elements of the Yukawa matrices $Y^{d,u,e,\nu}$ are subject to conditions on $U(1)'$ charges, discussed in \Sec{YukInv}.

The scalar potential consists of mass and cubic terms
\begin{equation}
\begin{aligned}
     V^{(2)} + V^{(3)}  & =  m_H^2 \,(H^\dagger H) + m_S^2 \, \tr(S^\dagger S)  \\
     & \phantom{ = }  + m_\phi^2  \,(\phi^\dagger \phi) + \mu_\text{det} \det\left(S + S^\dagger\right)  \,,
\end{aligned}
\end{equation}
 with the last term being a determinant in flavor space, as well as quartic interactions
\begin{equation}\label{eq:V4}
\begin{aligned}
    V^{(4)} &=  \lambda \,(H^\dagger H)^2 + s\,(\phi^\dagger \phi)^2  + u\,\tr(S^\dagger S S^\dagger S) \\
    &\phantom{=}  + v\, \tr(S^\dagger S) \tr(S^\dagger S) + \delta\,(H^\dagger H)\, \tr(S^\dagger S) \\
    &\phantom{=} + \dt\,(H^\dagger H)(\phi^\dagger \phi) + w\,(\phi^\dagger \phi)\, \tr(S^\dagger S)\,,
\end{aligned}
\end{equation}
featuring the Higgs $(\lambda)$ and BSM $(u,v,s)$ self-interactions 
as well as portal couplings $(\delta,\dt,w)$.
Further, two types of vacuum configurations exist in the BSM sector, a flavor-symmetric one $(V^+)$ with quartic $u>0$, and a symmetry-broken one $(V^-)$ with $u<0$ \cite{Hiller:2019mou,Hiller:2020fbu}.
Altogether, we find that 
the classical potential \eq{V4} is bounded from below provided that \cite{Paterson:1980fc,Litim:2015iea,PlanckSafeQuark,Kannike:2012pe}
\begin{equation} \label{eq:Vstab}
\begin{aligned}
        & \qquad  \lambda > 0, \qquad \qquad   \Delta > 0, \qquad \qquad  s > 0,\\
        &\delta' = \delta  + 2\sqrt{\lambda \Delta} > 0,\quad
        \dt' = \dt +2 \sqrt{\lambda s} > 0, \\
        &w'  = w +2 \sqrt{s \Delta} > 0, \\
        &2\sqrt{\lambda \Delta s} + \delta \sqrt{s} + \dt \sqrt{\Delta} + w \sqrt{\lambda} + \sqrt{\delta' \dt' w'} > 0\,,
\end{aligned}
\end{equation}
where the parameter
\begin{equation}\label{eq:Delta}
  \Delta = \begin{cases}
    \tfrac{u}3 + v > 0 & \text{ for } u > 0   \quad (V^+) \\
    u + v > 0 & \text{ for } u < 0   \quad (V^-)
  \end{cases}
\end{equation}
depends on the ground state for the BSM scalars.

\subsection{Anomaly Cancellation}\label{sec:ACC}

The fermionic $U(1)'$ charges $F_{X_i}$ are subject to constraints from gauge anomaly cancellation conditions (ACCs).
An excellent introduction to the subject is given in~\cite{Bilal:2008qx}. 
For recent phenomenological applications, see for instance~\cite{Ellis:2017nrp,Allanach:2018vjg,Rathsman:2019wyk,Costa:2019zzy}. 
Since BSM fermions $\psi$ are vector-like \eq{vector-like}, they do not contribute to anomalies.
The ACCs hence constrain only the $U(1)'$ charges of the SM fermions~\cite{Allanach:2018vjg} and the ones of the right-handed neutrinos. Overall we obtain six independent conditions
\begin{equation}\label{eq:ACCs}
\begin{aligned}
& \quad\sum_{i=1}^3 \, \left[2F_{Q_i} \!-\! F_{U_i} \!-\! F_{D_i} \right]=0 \,, \\
& \quad\sum_{i=1}^3 \, \left[3 F_{Q_i} \!+\!  F_{L_i} \right]=0 \,, \\
& \quad\sum_{i=1}^3 \, \left[F_{Q_i} \!+\!  3 F_{L_i} \!-\! 8 F_{U_i} \!-\! 2 F_{D_i} \!-\! 6 F_{E_i} \right]=0 \,,\\
&\quad\sum_{i=1}^3 \, \left[6 F_{Q_i} \!+\! 2 F_{L_i} \!-\! 3 F_{U_i} \!-\! 3 F_{D_i} \!-\! F_{E_i} \!-\! F_{\nu_i} \right]=0 \,,  \\
& \quad\sum_{i=1}^3 \,  \left[F^2_{Q_i} \!-\! F^2_{L_i} \!-\! 2 F^2_{U_i} \!+\!  F^2_{D_i} \!+\!  F^2_{E_i} \right]=0 \,,\\
& \quad\sum_{i=1}^3 \! \left[6 F^3_{Q_i} \!+\! 2 F^3_{L_i} \!-\! 3 F^3_{U_i} \!-\! 3 F^3_{D_i} \!-\! F^3_{E_i} \!-\! F^3_{\nu_i} \right]=0\,,
\end{aligned}
\end{equation}
which stem from anomaly cancellation in $SU(3)_C^2 \times U(1)^\prime$, $SU(2)_L^2 \times U(1)^\prime$, $U(1)_Y^2 \times U(1)^\prime$, gauge-gravity, $U(1)_Y \times U(1)^{\prime 2}$ and $U(1)^{\prime 3}$, respectively.
Since the right-handed neutrinos are singlets under the SM gauge group, their charges $F_{\nu_i}$ only appear in the $U(1)^{\prime 3}$ and gauge-gravity constraints in \eq{ACCs}. 

Overall, 18 charges are constrained by six ACCs, or 15 charges versus  six ACCs if the  right-handed neutrinos are  decoupled, with $F_{\nu_i} = 0$.
However, it turns out that combining the ACCs with the other theoretical and experimental constraints in this section -- in particular reproducing the $B$-anomalies -- 
requires $F_{\nu_i} \neq 0$ for most benchmarks, see also \Sec{Benchmarks} and App.\ref{sec:AppRHN}.
Note that three additional unconstrained $U(1)^\prime$ charges $F_{\psi,H,\phi}$ remain after taking \eq{vector-like} into account.

\subsection{Yukawa  Interactions and \texorpdfstring{$U(1)'$}{U(1)'} }
\label{sec:YukInv}

Here we discuss the implications of $U(1)'$ symmetry  for the Yukawa sector.
Note that the BSM Yukawa coupling $y$ is already assured to be gauge invariant by \eq{vector-like}.
On the other hand, requiring 
gauge invariance for the Yukawa couplings of quarks and leptons \eq{Yuk} would imply the following set of conditions 
\begin{equation}\label{eq:YCCs}
\begin{aligned}
    Y^u_{ij}: \qquad 0 &= F_{Q_i} + F_H - F_{U_j}, \\
    Y^d_{ij}: \qquad 0 &= F_{Q_i} - F_H - F_{D_j}, \\
    Y^e_{ij}: \qquad 0 &= F_{L_i} - F_H - F_{E_j}, \\
    Y^\nu_{ij}: \qquad 0 &= F_{L_i} + F_H - F_{\nu_j},
\end{aligned}
\end{equation}
for $i,j=1,2,3$.
A realistic model with CKM and PMNS mixing requires to switch on  off-diagonal entries, at least for the up- or the down-sector. Further inspection then
generically yields {\it universality}, that is, $F_{Q_1}=F_{Q_2}=F_{Q_3}$ and similarly for the other matter representations.
As a result, no tree-level FCNC is  induced at the $Z^\prime$-$b$-$s$-quark vertex,  see App.\ref{sec:chargeMixDetails}, prohibiting this  solution to the $B$-anomalies.

Here we instead pursue a model set-up that is {\it i)} 
radiatively stable under RG-evolution including the dominant Yukawa contributions,  {\it ii)}  reaches the Planck scale safely and  {\it iii)}  explains the $B$-anomalies. 
To do so, we  impose  \eq{YCCs} for  the diagonal quark Yukawa elements only, allowing for corresponding mass terms
\begin{equation}\label{eq:QuarkMassYCCs}
\begin{aligned}
    Y^u_{ii}: \qquad 0 &= F_{Q_i} + F_H - F_{U_i}\,, \\
    Y^d_{ii}: \qquad 0 &= F_{Q_i} - F_H - F_{D_i}\,. \\
\end{aligned}
\end{equation}
A few comments are in order:
Firstly, we do not impose  conditions for the lepton sector, as in this work
lepton and neutrino masses are neglected: they are numerically negligible for the RG-analysis at hand and 
also decouple from the remaining beta-functions.\footnote{In addition, $U(1)'$-gauge invariance of $Y^{e,\nu}_{ii}$ in combination with the other constraints discussed in  this section  yields $F_{L_2} = F_{E_2}$ and thus $\Cten=0$, see \Sec{B-Anomalies} for details. As we are however also interested in BSM scenarios with $\Cten \neq 0$ we refrain from using the  $Y^{e,\nu}_{ii}$ condition.}
As such,  terms in $Y^{e,\nu}_{ij}$ that are accidentally allowed in specific benchmarks, and potential LFV effects, become immaterial.
Secondly,  \eq{QuarkMassYCCs}  does not entirely reflect the hierarchy of all SM Yukawa elements.
While only the top coupling $Y^u_{33}$ is truly  essential for the RG evolution, other Yukawas remain naturally orders of magnitude smaller under the RG running.
A more minimal model set-up could be envisaged in which only the Yukawas for the bottom and the top are switched on via  \eq{QuarkMassYCCs}.
This setting is also radiatively stable und captures the dominant Yukawa contributions, leading to identical benchmarks as the three-generation case.
We comment on structural differences in App.\ref{sec:AppRHN}.
We finally stress that  addressing the flavor puzzle, for instance by promoting the $U(1)'$ to a Froggatt-Nielsen symmetry to obtain flavor patterns, requires additional Higgs doublets {\it e.g.}\cite{Correia:2019pnn}, an endeavor which is  beyond the scope of this work.

\subsection{Gauge-Kinetic Mixing}\label{sec:KinMix}

In the gauge sector, kinetic mixing occurs between the abelian vector fields via the parameter $\eta$ with $|\eta| < 1$ and
\begin{equation}\label{eq:kin-mix}
    \mathcal{L}_\text{abel} = \frac{-1}{4(1 - \eta^2)} \begin{pmatrix} F^{\mu\nu} \\[.1em]  F'^{\mu\nu} \end{pmatrix}^{\intercal} \!\begin{pmatrix} 1 & -\eta \\[.2em]  -\eta & 1 \end{pmatrix} \begin{pmatrix} F_{\mu\nu} \\[.2em]   F'_{\mu\nu} \end{pmatrix} \,,
\end{equation}
where $F_{\mu\nu}$ and $F'_{\mu\nu}$ denote the field strength tensors of the $U(1)_Y$ and $U(1)'$ interactions, respectively. 
Note that $\eta$ is not natural and cannot be switched off adjusting theory parameters. 
Moreover, it violates custodial symmetry and introduces mass mixing between $Z$ and $Z'$ after electroweak symmetry breaking.
Additional mixing terms are also generated if the SM Higgs is involved in the $U(1)'$ breaking. This is avoided by fixing
\begin{equation}\label{eq:FH=0}
    F_H = 0,
\end{equation}
which ensures that the scales and mechanisms of electroweak symmetry and $U(1)'$ breaking are independent, as these phenomena are solely mediated via $H$ and $\phi$, respectively.
This implies that only $\eta$ yields corrections to the $\rho$ parameter, $\rho = M_W/(M_Z \cos \theta_W)$, at tree-level
\begin{equation}
    \rho^{-1} = 1 + \frac{\eta^2 \sin^2\theta_W}{1 - z^2} \quad  \text{with} \quad z = \left(\frac{M_Z}{M_{Z'}}\right)\Big|_{\eta = 0}\,.
\end{equation}
However, a global fit of electroweak precision parameters in \cite{Zyla:2020zbs} suggests a new physics (NP) contribution
 \begin{equation}\label{eq:rho-NP}
    \left(\frac{\delta \rho}{\rho}\right)^\text{NP} = (3.8 \pm 2.0) \cdot 10^{-4}\,.
\end{equation}
 of the opposite sign. 
 Hence, barring any cancellations from other sources, one would expect the kinetic mixing to be subleading at the electroweak scale, which roughly requires
 \begin{equation}\label{eq:eta_bound}
 |\eta| \ \lesssim \ \mathcal{O}(10^{-2})\,.
  \end{equation}
 Furthermore, SM fermions $X_i$ attain corrections to their photon and $Z$ couplings $g_{X_i}^{\gamma,Z} \propto \eta \,F_{X_i}\, g_4$, 
 which can be evaded for small values of the $U(1)'$ gauge coupling $g_4$ or $\eta$ at the electroweak scale.

\subsection{Scalar Symmetry Breaking and Mixing \label{sec:SSB}}

To generate a $Z'$ mass, we consider spontaneous breaking of the $U(1)'$ symmetry by the vacuum expectation value
$\langle\phi\rangle = v_\varphi/\sqrt{2} \neq 0$, which yields
\begin{equation} \label{eq:MZprime}
    M_{Z'} = |F_\phi|\,g_4\,v_\varphi\,,
\end{equation}
while there is no contribution from the SM Higgs $H$ owning to \eq{FH=0}.
After electroweak symmetry breaking
\begin{equation}
    H = \frac{1}{\sqrt{2}} \begin{pmatrix} 0 \\ v_h + h \end{pmatrix}, \qquad \phi = \frac{1}{\sqrt{2}}(v_\varphi + \varphi),
\end{equation}
only the real modes $h$ and $\varphi$ survive, which are rotated into mass eigenstates
\begin{equation}
     \begin{pmatrix} h' \\ \varphi' \end{pmatrix} = \begin{pmatrix} \cos \beta & -\sin \beta  \\ \sin \beta & \cos \beta \end{pmatrix} \begin{pmatrix} h \\ \varphi \end{pmatrix}
\end{equation}
via the mixing angle
\begin{equation}
    \tan 2 \beta = \frac{ \widetilde{\delta} \,v_h \,v_\varphi}{s \,v_\varphi^2 - \lambda \, v_h^2} \approx \frac{\widetilde{\delta}}{s} \frac{v_h}{v_\varphi}
    \label{eq:beta}
\end{equation}
induced by scalar portal couplings. 
The scalar mixing opens many decay channels of $\varphi'$ to SM fermions or gauge bosons. 
On the other hand, the decay width of $h'$ to SM final states $\{f\}$ is reduced\footnote{Note that $f = ZZ^*$ also receives contributions from gauge-kinetic mixing.} via a global factor
\begin{equation}
\Gamma(h' \rightarrow \{f\})= \cos^2 \beta \; \Gamma^\text{SM}(h \rightarrow \{f\})\,.
\label{eq:Higgswidth}
\end{equation}
This implies a constraint
\begin{equation} \label{eq:betabound}
    \sin^2 \beta \leq 0.01 
\end{equation}
due to combined Higgs signal strength measurements from several channels \cite{Zyla:2020zbs}. 
Combining \eq{MZprime}, \eq{beta} and \eq{betabound} yields roughly
\begin{equation} \label{eq:scalarMix}
    |F_\phi|\, g_4 \frac{\dt}{s} < \begin{cases}
4.1 & \text{ for } M_{Z^\prime}=5 \,\TeV\\
2.4 & \text{ for } M_{Z^\prime}=3 \,\TeV
\end{cases}
\end{equation}
which can be easily accomplished as $|F_\phi| g_4 = \mathcal{O}(1)$.

The scalar $S$ may also develop a non-vanishing vacuum-expectation value, as considered in \cite{Hiller:2020fbu}. 
This leads to a more complicated mixing into mass eigenstates that is however controlled by the size of the scalar portal couplings $|\delta|$, $|\dt|$ and $|w|$.

\section{\bf Explaining the \texorpdfstring{$B$}{B}-anomalies}
\label{sec:B-Anomalies}

In this section we give the weak effective theory description of $b \to s \ell^+ \ell^-$ transitions, which allows to perform a global fit
to $b \to s$ data (Sec.~\ref{sec:EFT}), to match the $Z^\prime$ model (Sec.~\ref{sec:Zpmatch}) and to identify
viable benchmarks (Sec.~\ref{sec:Benchmarks}).

\subsection{EFT Description and Fits \label{sec:EFT}}

The effective hamiltonian for $b \to s \ell^+ \ell^-$ transitions can be written as  \cite{Aebischer:2019mlg,Kriewald:2021hfc}
\begin{equation} \label{eq:Heff}
\begin{aligned}
    \mathcal{H}_\text{eff}^{bs\ell\ell} \supset & -\frac{4\,G_F}{\sqrt{2}}\frac{\alphae}{4\pi}V_{tb}\,V_{ts}^\ast\, \\
    & \sum_{\ell=e,\mu,\tau} \sum_{i=9,10} \left( c_i^{\ell}\,\mathcal{O}_i^{\ell} + c_i^{\prime \ell}\,\mathcal{O}_i^{\prime \ell}\right)\,,
\end{aligned}
\end{equation}
with the semileptonic, dimension six operators
\begin{equation}\label{eq:O9O10}
\begin{aligned}
    \mathcal{O}_{9\phantom{0}}^{\ell} =& \left( \bar{s}_L\gamma_\mu b_L\right)\left( \bar{\ell}\gamma^\mu \ell \right),\\
    \mathcal{O}_{10}^{\ell} =& \left( \bar{s}_L\gamma_\mu b_L\right)\left( \bar{\ell}\gamma^\mu\gamma^5 \ell \right)\,,
\end{aligned}
\end{equation}
where the primed operators $\mathcal{O}_i^{\prime \ell}$ are obtained by interchanging the quark chiralities $L \leftrightarrow R$, and $\alpha_e$ ($G_F$) denotes the fine structure (Fermi's) constant.
The Wilson coefficients $c_i^{\ell}$ include a lepton universal SM contribution, $C_i^\text{SM}$, and a pure NP one, $C_i^{\ell}$, that is, 
$c_i^{\ell}=C_i^\text{SM}+C_i^{\ell}$. 
Note that we do not consider lepton flavor violating new physics contributions, which is reflected in the ansatz \eq{O9O10}.

\begin{figure}
    \centering
    \includegraphics[width=.49\textwidth]{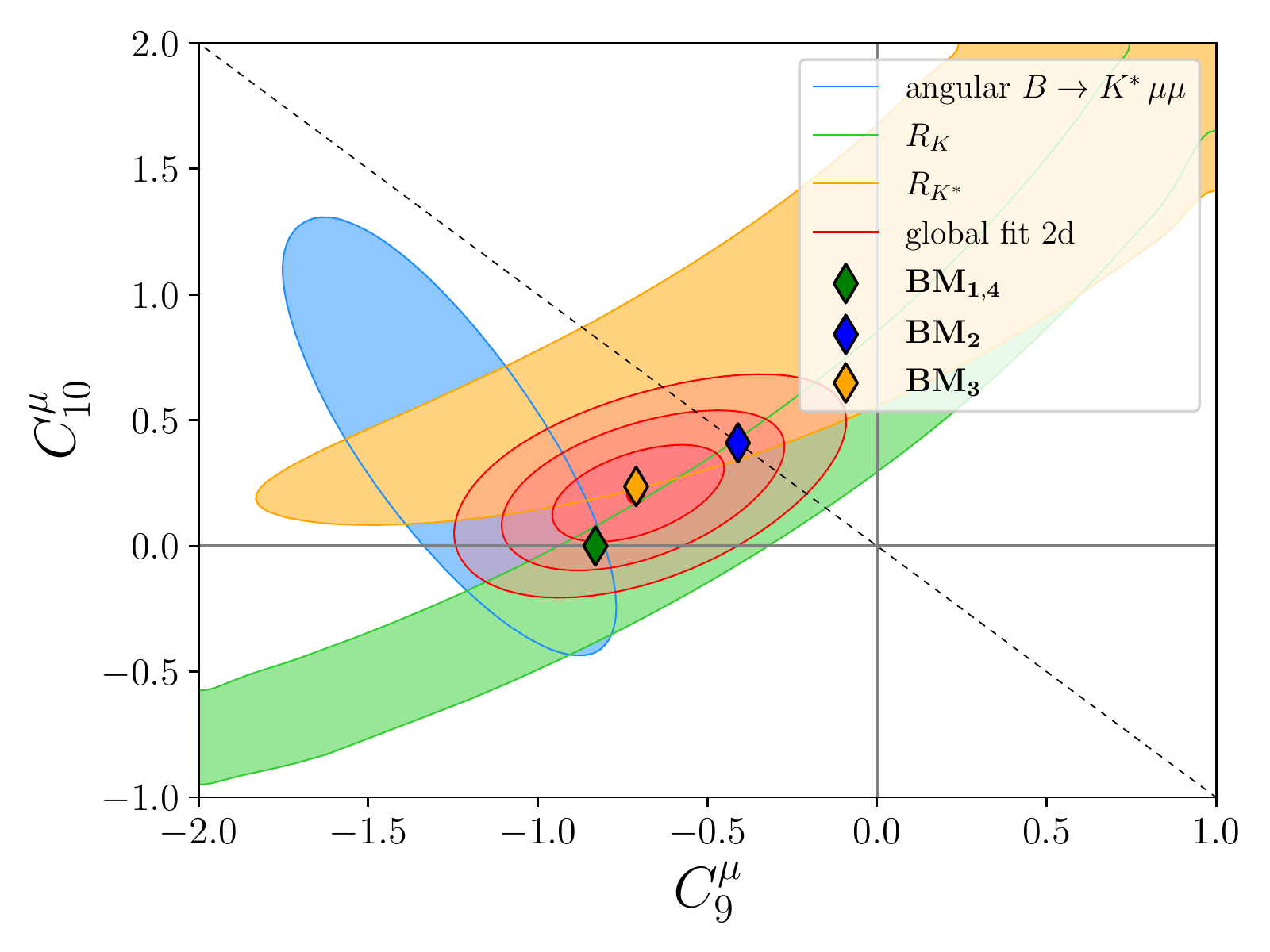}
    \caption{Likelihood contours from the $B$-anomalies using the 2d fit results from \tab{FitValues} in the plane of new physics Wilson coefficients $\Cnine$ vs. $\Cten$. 
    The red contours (at 1-3$\,\sigma$) incorporate all observables used in the fit, 
    whereas we also show contours (at $1\,\sigma$) using a subset of these observables, e.g. only $R_K$ (green), only $R_{K^\ast}$ (yellow) and $B \to K^\ast \mu^+\mu^-$ angular observables (blue).
    The four benchmarks \eq{benchmarks} are displayed as diamond-shaped symbols, the dashed black line indicates $\Cnine=-\Cten$. 
    Note that \BM{1} and  \BM{4}   generate identical values of $\Cnt$.
    See the main text for further details.
    }
    \label{fig:c9c10_rk_benchmarks}
\end{figure}

We use the results of a global fit of Wilson coefficients to $b \to s $ data  of Ref.~\cite{Bause:2021ply}, which have been performed using the tool \texttt{flavio}~\cite{Straub:2018kue} with several $b\rightarrow s \ell^+ \ell^-$ observables including the most recent LHCb measurement of $R_K$~\cite{Aaij:2021vac}. 
These observables incorporate (binned) branching ratios, angular observables such as $A_{FB}$, $F_H$ and $P_i^{(\prime)}$, as well as $R_{K^{(\ast)}}$, where results from Belle and LHCb measurements in selected low and high-$q^2$ bins have been employed.
For further details on the fit input as well as the results we refer to Ref.~\cite{Bause:2021ply}.
An overview of best fit results from \cite{Bause:2021ply} for the Wilson coefficients $\Cntprnopr$ in different new physics scenarios corresponding to 1d, 2d and 4d fits is given in \tab{FitValues}.
\fig{c9c10_rk_benchmarks} displays the likelihood contours in the $\Cnine-\Cten$ plane using the global fit results for the 2d scenario (red) of \tab{FitValues} as well as contours for different subsets of observables to emphasize the impact of $R_{K^{(\ast)}}$ in these fits. 

Large pulls $\gtrsim 6\, \sigma$ are consistent with \cite{Kriewald:2021hfc} and give strong support for  the $B$-anomalies.
We also observe that presently contributions to $\mathcal{O}_i^{\prime \ell}$ are not needed, and can therefore be neglected. 
We return to a discussion of right-handed currents in view of possible future data in \Sec{RH-fcnc}.
We also note that right-handed currents can alleviate possible tensions with $B$-mixing, see \App{AppBSMixing}.

\begin{table}
\def\arraystretch{1.5}
\centering
\setlength\arrayrulewidth{1.3pt}
 \setlength{\tabcolsep}{1pt}
\resizebox{0.48\textwidth}{!}{
\begin{tabular}{|c|c|c|c|c|c|c|}
\hline
Dim. & Fit for & $\Cnine$ & $\Cten$ & $\Cninepr$ & $\Ctenpr$ & $\text{Pull}_ \text{SM}$ \\ \hline
1d & $\Cnine$ & $-0.83 \pm 0.14$ & - & - & - & $6.0 \, \sigma$ \\ \hline 
1d & $\Cnine=-\Cten$ & $-0.41 \pm 0.07$ &  $-\Cnine$ & - & - & $6.0 \, \sigma$\\ \hline 
2d & $\Cnt$ & $-0.71 \pm 0.17$ & $0.20 \pm 0.13$ & - & - & $5.9 \, \sigma$\\ \hline 
4d & $\Cntprnopr$ & $-1.07 \pm 0.17$ & $0.18 \pm 0.15$ & $0.27 \pm 0.32$ & $-0.28 \pm 0.19$ & $6.5 \, \sigma$ \\ \hline
\end{tabular}
}
\caption{Best fit values for the Wilson coefficients $\Cntprnopr$ in different new physics scenarios and their respective pull from the SM hypothesis. Table entries taken from \cite{Bause:2021ply} (see there for further details).
} 
\label{tab:FitValues}
\end{table}

\subsection{Matching \texorpdfstring{$Z^\prime$}{ZPr} Models  \label{sec:Zpmatch}}

\begin{figure}
    \centering
	\includegraphics[scale=1.2]{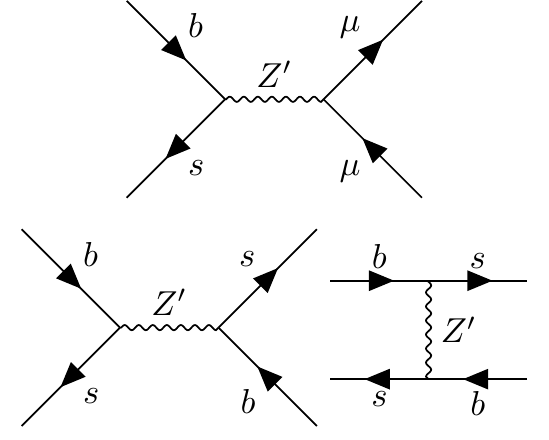}
	\caption{Top: Tree-level $b \to s \mu \mu$ transition in an $U(1)^\prime$ extension of the SM. Bottom: $Z'$ contributions to $B_s$-mixing.}
    \label{fig:BSMumu_treelevel}
\end{figure}

In the $Z'$ models presented here, the Wilson coefficients are generated by tree-level contributions of a $Z^\prime$ gauge boson coupling to $b,s$-quarks as well as muons as depicted in \fig{BSMumu_treelevel}.
The requisite $Z^\prime$ couplings  can be written as
\begin{align}
 \begin{split}
    \mathcal{L}_{Z^\prime} \supset \,& \left(
    g_L^{bs}\,\bar s_L \gamma^\mu b_L Z^\prime_\mu + g_R^{bs}\,\bar s_R \gamma^\mu b_R Z^\prime_\mu +\text{h.c.} \right) \\
    & \quad + \, g_L^{\ell\ell}\bar \ell_L \gamma^\mu \ell_L Z^\prime_\mu+g_R^{\ell\ell}\bar \ell_R \gamma^\mu \ell_R Z^\prime_\mu  \,,
\end{split}   \label{eq:LintZ}   
\end{align}
where $\ell=e,\mu,\tau$. 
Integrating out the $Z^\prime$ at tree-level, one obtains the effective Hamiltonian (cf. \cite{DiLuzio:2019jyq})
\begin{equation}\label{eq:effHZprime}
 \begin{aligned}
    \mathcal{H}_{Z^\prime}^\text{eff} \supset \frac{1}{2M_{Z'}^2}& \Big[ 
    \phantom{\ + }g_L^{bs} (\overline{s}_L \gamma^\mu b_L) + g_L^{bs*} (\overline{b}_L \gamma^\mu s_L) \\
    &\ + g_R^{bs} (\overline{s}_R \gamma^\mu b_R) + g_R^{bs*} (\overline{b}_R \gamma^\mu s_R) \\
    &\ + g_L^{\ell\ell}(\bar{\ell}_L \gamma^\mu \ell_L) \: + g_R^{\ell\ell}(\bar{\ell}_R \gamma^\mu \ell_R)\ \Big]^2,
 \end{aligned}
\end{equation}
which gives rise to contributions  to semileptonic coefficients $C_{9,10}^{(\prime)\ell}$, as well as  $B_s$-mixing.

Comparing \eq{effHZprime} with \eq{Heff}, the NP Wilson coefficients  for $b \to s \ell^+ \ell^-$ transitions can be read off
\begin{align}\label{eq:WCs}
\begin{split}
 C_{9,10}^{\ell} &= \mathcal{N}^{-1}\, \frac{g_L^{bs} \bigl( g_R^{\ell\ell} \pm g_L^{\ell\ell} \bigr)}{M_{Z^\prime}^2} \,,\\ 
C_{9,10}^{\prime\,\ell} &= \mathcal{N}^{-1}\,\frac{g_R^{bs} \bigl( g_R^{\ell\ell} \pm g_L^{\ell\ell} \bigr)}{M_{Z^\prime}^2} \,,   
\end{split}
\end{align}
with
\begin{align}
    \mathcal{N}= -\frac{\sqrt{2}\,G_F\,\alphae}{\pi}\,V_{tb}\,V_{ts}^\ast \,.
\end{align}

The flavor diagonal couplings of the $Z^\prime$ to leptons, $g_{L(R)}^{\ell\ell}$, are given by 
$(i=1,2,3 \leftrightarrow \ell = e,\mu,\tau)$
\begin{equation}\label{eq:gll}
g_{L(R)}^{\ell\ell} = g_4\,F_{L_i(E_i)} \, . 
\end{equation}
The non-diagonal left-handed $bs$-quark couplings read
\begin{align}\label{eq:gbs}
    g^{bs}_L &=  V_{tb}\,V_{ts}^\ast\left(F_{Q_3}-F_{Q_2}\right) \cdot g_4 \,, 
\end{align}
with $V_{tb}\,V_{ts}^* \approx -0.04$  \cite{Zyla:2020zbs}.
Couplings to right-handed FCNCs can always be switched off by a suitable rotation between right-chiral down-type flavor and mass eigenstates, such that
\begin{equation} \label{eq:gRbq}
    g^{bs}_R = 0 \, , 
\end{equation}
hence $\Cntpr = 0$,
which is sufficient to explain present data, see \tab{FitValues}.
 Note that we discard the possibility of large cancellations between up- and down-quark flavor rotations.
Details on flavor rotations are given in \App{chargeMixDetails}.

We learn that NP scenarios with $\Cnine=-\Cten$ require $g_R^{\mu\mu}=0$, that is, $F_{E_2}=0$,
whereas models with $\Cten=0$ require vector-like muon charges $F_{L_2}=F_{E_2}$.

The $2d$ fit results shown in \tab{FitValues} exhibit the pattern
\begin{equation} \label{eq:C9-C10-pattern}
-\Cnine  \geq  \Cten \geq 0\, , 
\end{equation}
which has to be matched by the $U(1)^\prime$ models.
Thus, \eq{C9-C10-pattern} in combination with $g_4^2>0$ as well as the known signs of the relevant CKM elements implies conditions for the $U(1)'$ charges. 
We find the following constraints on the  $U(1)^\prime$ charge assignments
\begin{equation} \label{eq:BAnomaliesChargeConstr}
\begin{aligned}
& (F_{Q_3} > F_{Q_2} \quad \text{and} \quad F_{L_2} \geq F_{E_2} \geq 0) \quad\\
\text{or} \quad &(F_{Q_3} < F_{Q_2} \quad \text{and} \quad F_{L_2} \leq F_{E_2} \leq 0) \, .
\end{aligned}
\end{equation}

A $Z'$-coupling  $g_L^{bs} \neq 0$ (see \eq{LintZ})  also invariably generates a tree-level contribution to $\Delta M_s$, i.e. to $B_s$-mixing, see \fig{BSMumu_treelevel}.
For $g_R^{bs}=0$, this implies
\begin{align}\label{eq:Bmix_gL_bound}
    \frac{\left|g_L^{bs}\right|^2}{M_{Z^\prime}^2} \lesssim 
    1.24 \cdot 10^{-5} \text{ TeV}^{-2}
\end{align}
at 99\% c.l.~\cite{Dwivedi:2019uqd}. 
While this bound is quite strong, 
it can be fulfilled by the benchmark models as shown in the next section \ref{sec:Benchmarks}.

\subsection{Benchmarks}
\label{sec:Benchmarks}

\begin{table*}[t!]
\setlength\arrayrulewidth{1.3pt}
\def\arraystretch{1.5}
 \setlength{\tabcolsep}{1pt}
    \centering
	\begin{tabular}{|c|ccc|ccc|ccc|ccc|ccc|ccc|c|c|c|}
	\hline
	Model & & $F_{Q_i}$& & & $F_{U_i}$ & & & $F_{D_i}$&  & & $F_{L_i}$&  & & $F_{E_i}$ & & & $F_{\nu_i}$&  & $F_H$ & $F_\psi$ & $F_\phi$\\
	\hline
	\BM1 & $  \phaMinus \frac{1}{20}$ &   $\phaMinus \frac{1}{20}$ & $-\frac{1}{10} $ & $ \phaMinus \frac{1}{20}$ & $\phaMinus \frac{1}{20} $ & $-\frac{1}{10}$ & $\phaMinus \frac{1}{20}$ & $\phaMinus \frac{1}{20} $ & $-\frac{1}{10}  $ & $ \phaMinus 0 $ & $-\frac{9}{10}$ & $\phaMinus \frac{9}{10}$ &
	$ \phaMinus 0$ & $-\frac{9}{10}$ & $\phaMinus \frac{9}{10} $ & 
	$ \phaMinus 0$ & $\phaMinus 0$ & $\phaMinus 0$ & 
	$0$ & $1$ & $\frac{1}{5}$ \\
	\BM 2 & $ -\tfrac14 $ & $-\tfrac14 $ & $\phaMinus \tfrac16 $ &
	$  -\tfrac14 $ &  $-\tfrac14 $ & $\phaMinus \tfrac16 $ & 
	$  -\tfrac14 $ &  $-\tfrac14 $ & $\phaMinus \tfrac16 $ & 
	$ \phaMinus 0$ & $\phaMinus 1$ & $\phaMinus 0 $ & 
	$ \phaMinus 0$ & $\phaMinus 0$ & $\phaMinus 1 $ & 
	$ \phaMinus \frac{1}{12}$ & $-\frac{1}{12}$ & $\phaMinus 1 $ & 
	$0$ & $\frac{11}{12}$ & $\tfrac19$ \\
	\BM 3 &$ -\tfrac18$ & $-\tfrac18$ & $\phaMinus 0  $ & 
	$ -\tfrac18$ & $-\tfrac18$ & $\phaMinus 0 $ & 
	$ -\tfrac18$ & $-\tfrac18$ & $\phaMinus 0 $ & 
	$ \phaMinus 0 $ & $\phaMinus \tfrac12 $ & $\phaMinus \tfrac14 $ & 
	$ \phaMinus 0 $ & $\phaMinus \tfrac14 $ & $\phaMinus \tfrac12 $ & 
	$ \phaMinus 0 $ & $\phaMinus \tfrac14 $ & $\phaMinus \tfrac12 $ & 
	$0$ & $1$ & $\tfrac18$ \\	
	\BM 4 & $\phaMinus 0$ & $\phaMinus 0$ & $\phaMinus\frac{1}{9}$ & $\phaMinus 0$ & $\phaMinus 0$ & $\phaMinus\frac{1}{9}$ & $\phaMinus 0$ & $\phaMinus 0$ & $\phaMinus\frac{1}{9}$ & 
	$\phaMinus0$ & $\phaMinus\frac{1}{3}$ & $-\frac{2}{3}$ & 
	$\phaMinus0$ & $\phaMinus\frac{1}{3}$ & $-\frac{2}{3}$ & 
	$\phaMinus0$ & $\phaMinus\frac{1}{3}$ & $-\frac{2}{3}$ & 
	$0$ & $1$ & $\tfrac16$ \\		
	\hline
	\end{tabular}
	\caption{$U(1)^\prime$ charge assignments $F_X, \, X=Q_i, U_i, D_i, L_i, E_i, \psi, H, \phi$ in the four benchmark models.}
	\label{tab:Benchmarks}
\end{table*}

In this section we identify three benchmark models (BMs) predicting new physics around a matching scale of $\mu_0 \simeq 5$~TeV, 
as well as a fourth scenario where the scale of new physics can be lower $\mu_0 \simeq 3$~TeV. 
Explicit $U(1)'$ charge assignments are given in \tab{Benchmarks}.
The BMs induce different patterns in the semileptonic Wilson coefficients relevant for explaining the $B$-anomalies:
\begin{equation}\label{eq:benchmarks}
\begin{aligned}
\qquad \textbf{ \BM {1,4}\,: } \qquad &\phantom{-\ }\Cnine \neq 0 \ \text{ and } \ \Cten = 0\,, \\
\qquad \textbf{ \BM {2\phantom{,4}}\,: } \qquad &\phantom{-\ }\Cnine = -\Cten\,, \\
\qquad \textbf{ \BM {3\phantom{,4}}\,: } \qquad &-\Cnine \gg \Cten > 0\, \\
\end{aligned}
\end{equation}
The corresponding Wilson coefficients in the $\Cnine$-$\Cten$-plane are depicted in \fig{c9c10_rk_benchmarks}.
Apart from accounting for $R_{K^{(*)}}$ and other observables in the fit \tab{FitValues}, these models are also compliant with the theoretical and phenomenological constraints from \Sec{Models}, in particular ACCs \eq{ACCs}, gauge invariance of quark mass terms \eq{QuarkMassYCCs}, $B_s$-mixing \eq{Bmix_gL_bound}, kaon mixing \eq{KaonBound} as well as electron electroweak precision measurements \eq{ElectronBound}. Note that the combination of all these constraints in \BM{2,3} invariably requires right-handed neutrinos with at least one $F_{\nu_i}\neq 0$,
see \App{AppRHN} for further details.
In particular, within our $Z^\prime$ model framework, $F_{\nu_i}=0$ necessarily implies $C_{10}=0$. This is indeed allowed by data,  but by no means a unique solution to the 
global fit, see \tab{FitValues}.
The benchmarks in this works share the feature $F_H=0$. Due to this simplification, the $Z'$ mass is generated independently from electroweak symmetry breaking.

Let us now discuss the specifics for each benchmark model in more detail.
\BM 1 is part of a family of theories considering a $U(1)'$ charge that is the same  for all representations of quarks $(q_i = Q_i$, $U_i$, $D_i)$ and leptons $(\ell_i= L_i$, $E_i)$ for a given generation $i$ with
\begin{equation}
    F_{q_3} = -2\,F_{q_2} = -2 F_{q_1}, \quad F_{\ell_1} = 0, \quad F_{\ell_3} = - F_{\ell_2}.
\end{equation}
Thus,  $\Cten = 0$ due to \eq{WCs}. 
In contrast to the other benchmarks, right-handed neutrinos are not necessarily required and hence omitted in \BM1.
In addition to quark masses, the particular choice of $U(1)'$ charges in \BM1 (see \tab{Benchmarks}) also allows for diagonal lepton  Yukawa elements $Y^{e}_{ii}$.
In order to reproduce the best fit value of $\Cnine=-0.83$ (cf. \tab{FitValues}) we obtain the matching condition $\alpha_4(\mu_0) = 1.87 \cdot 10^{-2}$
for the $U(1)'$ gauge coupling.

For \BM 2 we set $F_{E_2} = 0$ to induce the best fit values $\Cnine =-\Cten = -0.41$ (cf. \tab{FitValues}). This leads to
    $\alpha_4(\mu_0) = 5.97 \cdot 10^{-3}$
and 
\begin{equation}
    \frac{|g_L^{bs}|^2}{M^2_{Z^\prime}} = 1.16 \cdot 10^{-5} \; \text{TeV}^{-2}\,,
\end{equation}
near but lower than the  $B_s$-mixing bound \eq{Bmix_gL_bound}.
Possible future tensions with improved $B_s$-mixing data can still be evaded 
by considering additional small couplings $g_R^{bs} = \frac{1}{X_s} g_L^{bs} \ll g_L^{bs}$ 
with $X_s \approx 10$ that cancel $Z^\prime$ contributions to the $B_s$-mixing, 
see \Sec{AppBSMixing} for details.

A third benchmark, \BM 3, aims at generating the hierarchy $-\Cnine \gg \Cten > 0$. 
Matching $\alpha_4(\mu_0) = 4.60 \cdot 10^{-2}$, one obtains 
\begin{equation}
        \Cnine = -0.71,\qquad \Cten = +0.24 \,,
\end{equation}
in excellent agreement with the fit results in \tab{FitValues}. 
Note that this is not a trivial achievement since the coupling  $\alpha_4(\mu_0)$ is adjusted to match two best fit values for the Wilson coefficients. 
\BM3 also features vanishing $U(1)^\prime$ charges  of third generation quarks; requisite $Z^\prime$ couplings to (mass eigenstate) $b$-quarks are
reintroduced after flavor rotations, see App.~\ref{sec:chargeMixDetails}.

Finally, in the fourth benchmark \BM4  all first and second generation quark charges vanish.
This allows a  lower matching scale than in \BM{1,2,3}, as constraints on the $Z^\prime$ from the LHC are evaded by suppressing its production.
\BM 4 can also be viewed as minimal in the sense that it requires fewer couplings than the other benchmarks  (see \tab{Benchmarks}).
The model shares 
$F_{L_2}=F_{E_2}$, hence $C_{10}^\mu=0$, with \BM1.
We use identical $\Cnine=-0.83$ but somewhat different $\alpha_4(\mu_0=3 \, \text{TeV}) = 2.45  \cdot 10^{-2}$ due to the lower matching scale than  in \BM1.\footnote{A model  with a similar minimal setup 
has been put forward in \cite{Allanach:2020kss}.
Unlike \BM4, however, it necessitates  a large tuning in flavor rotations, as CKM-like $bs$-mixing  would violate \eq{BAnomaliesChargeConstr} and predict the wrong sign of $C_9^\mu$.}

\section{\bf  Planck Safety Analysis}
\label{sec:Planck-Safety}

Thus far, we have obtained phenomenologically viable $U(1)^\prime$ extensions that account for the $B$-anomalies. 
In this section we investigate conditions under which   the models are Planck-safe, by which we mean stable, well-defined, and predictive,  all the way up to the Planck scale $\MPl \approx 10^{19}$~GeV.

\subsection{Finding the BSM Critical Surface}
In order to identify Planck-safe  trajectories in the various benchmark models, we have to investigate the RG running of gauge, Yukawa, quartics and portal couplings
\begin{equation}\label{eq:couplings}
\begin{aligned}
\alpha_X& = \frac{X^2}{(4\pi)^2}\quad{\rm with}\quad X=\{g_1,g_2,g_3,g_4\}\,, \\
\alpha_Y &= \frac{Y^2}{(4\pi)^2}\quad{\rm with}\quad Y = \{y_t,\,y_b,\,y\}\,,\\
\alpha_Z &= \frac{Z}{(4\pi)^2}\quad {\rm with}\quad Z = \{\lambda, \, \delta, \, {\dt}, \, u, \, v, \, w, \, s\}\,,
\end{aligned}
\end{equation}
alongside the running of the kinetic mixing parameter $\eta$. 
The remaining Yukawa couplings of the SM can safely be neglected as they are numerically small and technically natural.
We then demand that the couplings \eq{couplings} remain finite and well-defined starting from the matching scale $\mu_0$  all the way up to the Planck scale where quantum gravity is expected to kick in. For the sake of this section, we set the matching scale to
\begin{equation}\label{eq:5TeV}
\mu_0 = \begin{cases}
5 \ {\rm TeV}& \text{ for \BM{1,2,3} } \\
3 \ {\rm TeV}& \text{ for \BM{4} }
\end{cases} \,.
\end{equation} 
For the numerical searches, we also assume that  all fields have masses at or below \eq{5TeV}, so that they can be treated  as effectively massless at energies above \eq{5TeV}. We shall further demand that the models display a viable ground state, corresponding to a stable scalar potential 
all the way up to $\MPl$ in compliance with  the stability conditions \eq{Vstab}.

On a practical level, then, we closely follow the bottom-up search strategy advocated in \cite{Hiller:2020fbu}, where the RG running of couplings is
evolved numerically from the matching scale up to the Planck regime. The stability of the quantum vacuum and the absence of poles is thereby checked along  any RG trajectory all the way up to $\MPl$.\footnote{In practice, we checked the absence of Landau poles and vacuum stability at 250 logarithmically equidistant points between the matching scale $\mu_0$ and the Planck scale $M_\text{Pl}$.}
For consistency, we adopt  RG equations at two loop accuracy for all couplings \cite{Machacek:1983tz,Machacek:1983fi,Machacek:1984zw,Luo:2002ti,Poole:2019kcm}, also using  \cite{Litim:2020jvl,Steudtner:FoRGEr} to extract perturbative beta functions from general expressions in the $\overline{\rm MS}$ scheme.\footnote{General  expressions  for $\overline{\rm MS}$ beta functions have recently been extended up to four, three, and two loops  in the gauge, Yukawa, and scalar couplings, respectively \cite{Bednyakov:2021qxa}. In the scalar-Yukawa sector $(\alpha_X=0)$,  general results for beta functions up to  three loops have been made available in \cite{Steudtner:2021fzs}.} All two-loop $\beta$-functions for our models are provided in an ancillary file.

For the matching to the SM, the values for SM couplings are extracted from experimental data at the electroweak scale \cite{Buttazzo:2013uya,Zyla:2020zbs,Athron:2017fvs} and subsequently RG evolved up to the matching scale \eq{5TeV}. Using the three-loop SM running of couplings \cite{Mihaila:2012pz,Bednyakov:2012rb,Bednyakov:2012en,Bednyakov:2014pia} and neglecting BSM threshold corrections, we find the matched couplings summarized in \tab{SM-match}.
\begin{table}[h]
    \setlength\arrayrulewidth{1.3pt}
\def\arraystretch{1.5}
    \centering
    \begin{tabular}{|c|cc|}
        \hline
         & $\ \ \mu_0 = 3\,\text{TeV} \ \  $ & $\ \  \mu_0 = 5\,\text{TeV} \ \ $ \\\hline
        $\alpha_1(\mu_0)$ & $8.40 \cdot 10^{-4}$ & $8.46 \cdot 10^{-4}$ \\
        $\alpha_2(\mu_0)$ & $2.54 \cdot 10^{-3}$ &  $2.52 \cdot 10^{-3}$ \\
        $\alpha_3(\mu_0)$ & $6.37 \cdot 10^{-3}$ & $6.09 \cdot 10^{-3}$ \\
        $\alpha_t(\mu_0)$ & $4.17 \cdot 10^{-3}$ & $4.00 \cdot 10^{-3}$ \\
        $\alpha_b(\mu_0)$ & $1.07 \cdot 10^{-6}$ & $1.02 \cdot 10^{-6}$ \\
        $\alpha_\lambda(\mu_0)$ & $5.18 \cdot 10^{-4}$ & $4.80 \cdot 10^{-4}$ \\ \hline
    \end{tabular}
    \caption{SM parameters at matching scales $\mu_0 = 3$~TeV and $\mu_0 = 5$~TeV.}
    \label{tab:SM-match}
\end{table}
In addition, the kinetic mixing parameter as well as the quartic coupling $\alpha_s(\mu_0)$ are constrained due to \eq{eta_bound} and \eq{scalarMix}, respectively, while $\alpha_4(\mu_0)$ is determined  by accounting for the $B$-anomalies in each of the benchmark models separately (\Sec{Benchmarks}). This leaves us with the values for the BSM Yukawa, quartic and portal couplings \begin{equation}\label{eq:free}
\left\{\alpha_{y,}, \alpha_{\delta},\, \alpha_{{\dt}},\, \alpha_u, \, \alpha_v, \, \alpha_w, \, \alpha_s \right\}\big|_{\mu=\mu_0}
\end{equation}
at the matching scale \eq{5TeV}  as  {\it prima facie} free model parameters.  The aim, then, is to identify the critical surface of parameters, by which we refer to the set of initial conditions which lead to well-defined RG trajectories up to Planckian energies.

\subsection{General Considerations}

\begin{figure}[b]
\centering
\includegraphics[trim=1.5cm 0 1.5cm 0, clip, width=0.9\columnwidth]{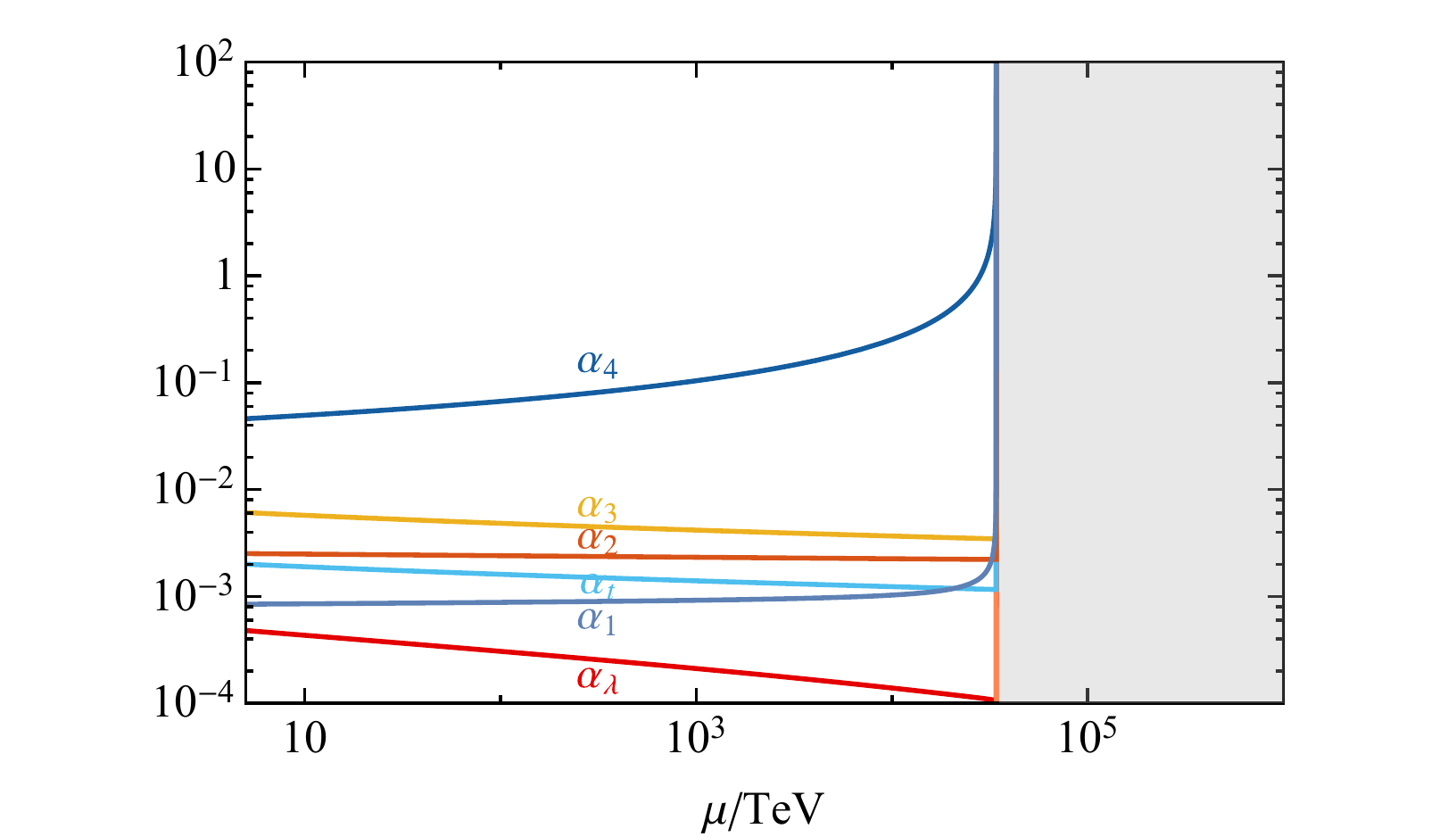}
\caption{The running of  gauge, top and Higgs couplings  in a  generic heavy $Z'$ extension explaining the $B$-anomalies shows a subplanckian Landau pole. 
The model in this plot has the same charges as shown for \BM3 in Tab.~\ref{tab:Benchmarks}, but {\it without} any of the BSM fields  $\psi,\,S$ and $\phi$ used in \BM3  (see main text).}
\label{fig:Pole}
\end{figure}

Before investigating  parameter constraints for concrete benchmarks, we elaborate on a few characteristics shared by all of them.
Explaining the $B$-anomalies typically requires $\alpha_4(\mu_0) \simeq \mathcal{O}(10^{-2})$, which implies a $U(1)'$ Landau pole around \eq{Landau} or even below. 
This result is illustrated, exemplarily, in \fig{Pole} which shows the  two-loop running of  the gauge, top, and Higgs couplings in a generic heavy $Z'$ model explaining the $B$-anomalies. Here, the $B$-anomalies require $\alpha_4(\mu_0=5\,\TeV)=4.60\cdot 10^{-2}$, see \eq{g4bound}, which leads to a Landau pole  around $3\cdot 10^4$~TeV. Notice that the pole arises much below the upper  bound \eq{Landau},
which is due to the fact that generic $U(1)'$ models cannot accommodate the strict minimal amount of $U(1)'$ charges required to satisfy the bound.

The addition of the $\psi$, $S$ and $\phi$ fields and their interactions 
has, in general, two  key effects. On one side, it lowers  the putative Landau pole due to additional $U(1)'$ charge carriers. 
On the other, it offers the prospect of moving putative poles past the Planck scale due to interactions.
Quantitatively, this necessitates compensating contributions through BSM Yukawas, typically of the order 
\begin{equation}\label{eq:boundYukawa}
F_\psi^2\,\alpha_{y}(\mu_0) \gtrsim \mathcal{O}(10^{-1})\,.
\end{equation}
At the same time, the absolute values of all other $U(1)'$ charges have to be sufficiently small relative to $|F_\psi|$ in order to help 
slow down the growth of $\alpha_4$. 
This is especially true  for the quarks and their charges, as they contribute to the growth of the $U(1)'$ gauge coupling with their color and isospin multiplicities. 
Incidentally, these considerations turn out to be in accord with the phenomenological constraints $|F_{Q_{2,3}}|/|F_\psi| \ll 1$ (cf. also \tab{Benchmarks}) which arises from $B_s$-mixing.
In turn, lepton charges have  to remain sizable enough to accommodate  the $B$-anomalies \eq{gll} in the first place. 

\begin{figure}[b]
\centering
\includegraphics[trim=1.5cm 0 1.5cm 0, clip, width=0.9\columnwidth]{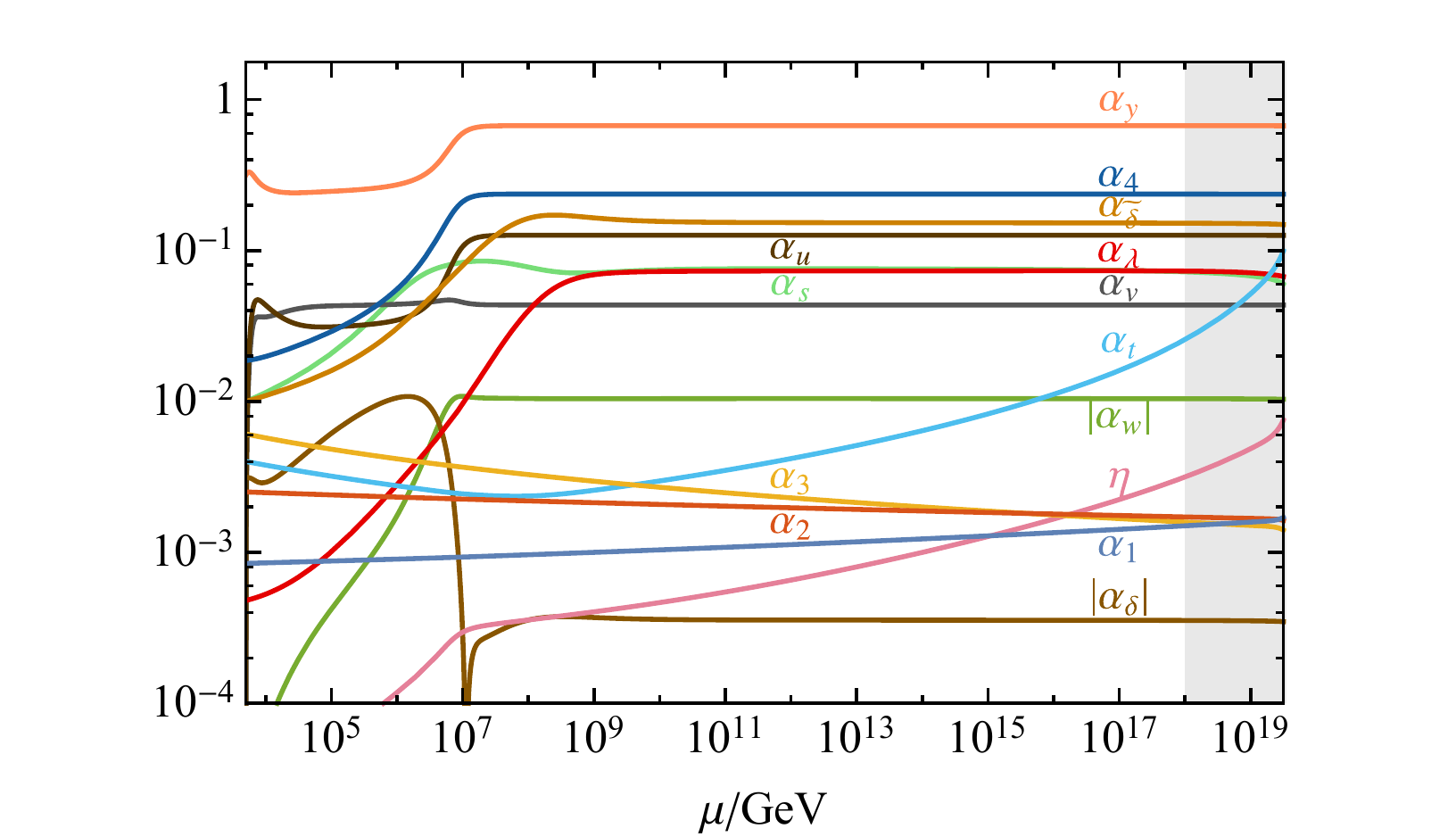}
\caption{Sample  running of couplings for \BM1 up to the Planck scale (gray area) showing trajectories for all couplings 
and kinetic mixing $\eta$
with initial conditions   
\eq{5TeV}, \tab{SM-match}  and
$\{\alpha_4, \eta,\, \alpha_{y,}, \alpha_{\delta},\, \alpha_{{\dt}},\, \alpha_u, \, \alpha_v, \, \alpha_w, \, \alpha_s \} |_{\mu_0} = \{1.87\cdot 10^{-2}\!,\, 0,\, 10^{-0.5}\!,\, 10^{-2.5}\!,\, 10^{-2}\!,\, 10^{-4}\!,\,  10^{-5}\!,\,  10^{-6}\!,\,  10^{-2}\}$.}
\label{fig:runBM1}
\end{figure}

Turning to the scalar sector, we observe  that the Higgs potential can only   be stabilized through  sizable contributions  from at least one of its portal couplings to the $S$ and $\phi$ scalars in \eq{V4}. Roughly speaking, we find that one, or the other, or both  should be  in the range
\begin{equation}\label{eq:range}
10^{-3} \lesssim \alpha_{\delta}(\mu_0),\alpha_{\dt}(\mu_0) \lesssim 10^{-1}\,.
\end{equation}
However, the required range of values for the  Yukawa coupling $\alpha_y$ may well destabilize the scalar sector if not countered by suitable contributions from the quartics and portals, and it would seem that vacuum stability and the absence of $U(1)'$ Landau poles  are mutually exclusive. 

Interestingly though, we find that both requirements can be met, the reason being that the slowing-down in the  $U(1)'$ sector entails a  slowing-down in the scalar sector \cite{Hiller:2020fbu}. Regimes with substantially slowed-down running of couplings are known as ``walking regimes'', and often relate to nearby fixed points in the plane of complexified couplings. Here, we observe that some or all beta functions in the extended scalar-Yukawa-$U(1)'$ sector may pass through a walking regime, whereby the growth of couplings comes nearly to a halt. Overall, this ensures that a subset of trajectories  can reach the Planck scale, after all.

\begin{figure}[b!]
    \centering
  \renewcommand*{\arraystretch}{0}
  \begin{tabular}{c}
    \includegraphics[width=.9\columnwidth]{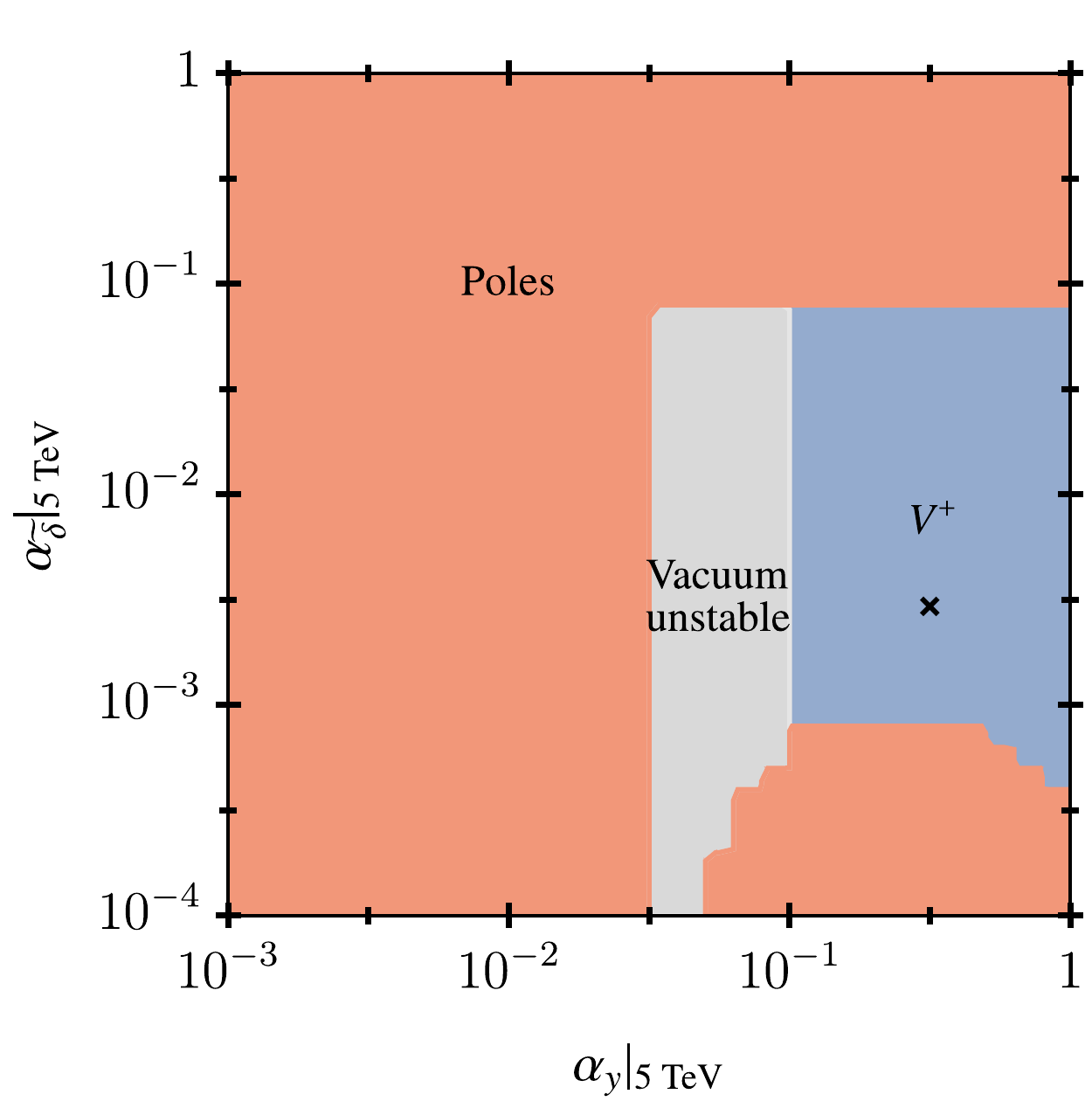}\\
    \includegraphics[width=.9\columnwidth]{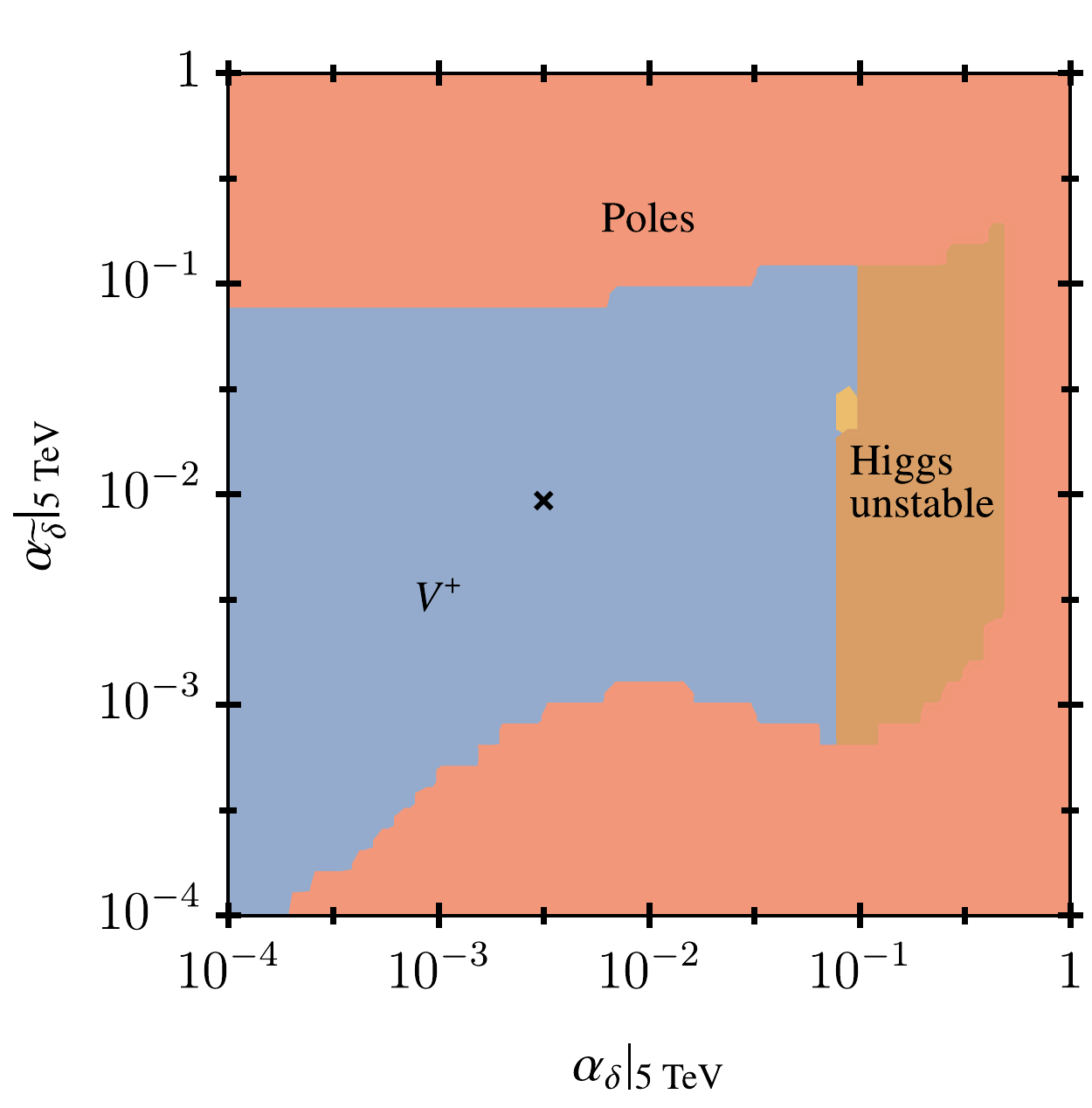}\\
  \end{tabular}
  \caption{Critical surface of parameters for  the benchmark model \BM1, projected onto the $\{\alpha_{y},\alpha_{\dt}\}|_{\mu_0}$ plane (top) and the $\{\alpha_{\delta},\alpha_{\dt}\}|_{\mu_0}$ plane (bottom). We further use the matching scale  \eq{5TeV}, together with  \tab{SM-match},
  $\alpha_\delta|_{\mu_0}=10^{-2.5}$ (top), $\alpha_y|_{\mu_0}=10^{-0.5}$ (bottom), and $\{\,\alpha_4,\, \eta, \,\alpha_u, \, \alpha_v, \, \alpha_w, \, \alpha_s \} |_{\mu_0}= \{1.87\cdot 10^{-2}\!,\, 0,\, 10^{-4}\!,\,  10^{-5}\!,\,  10^{-6}\!,\,  10^{-2}\}$. 
  The black cross corresponds to the trajectory in \fig{runBM1}. The color coding indicates  if the vacuum at the Planck scale is unstable (gray)  or stable with vacuum $V^+$ (blue), whether the Higgs is unstable (brown) or metastable (yellow), or whether poles have arisen (red).}
  \label{fig:ParameterSpaceBM1}
\end{figure}

As a final remark, we also find a considerable amount of parameter space where Planck-safe trajectories are enabled due to large gauge-kinetic mixing effects already at the matching scale, where  $|\eta(\mu_0)| \simeq \mathcal{O}(1)$ slows down the running of $\alpha_4$. 
Regrettably, however, this option is  excluded phenomenologically due to electroweak precision data \eq{eta_bound}.

In the remaining parts of this section, we perform a systematic scan over the free parameters \eq{free}
for each of the different benchmark models.

\subsection{Benchmark 1}

In the benchmark model \BM1 the  value required for the gauge coupling $\alpha_4$ to explain the $B$-anomalies comes out rather large,
$\alpha_4(\mu_0) = 1.87 \cdot 10^{-2}$. 
Hence switching-off the interactions of the BSM fields $\psi$ and $S$ would lead to a Landau pole around 110~TeV. 
We emphasize that the location of the na\"ive Landau pole  dictates an upper bound for the  masses of the $\psi$ and $S$ fields, $M_\psi, M_S<110$~TeV, as otherwise the theory would fall apart before these fields ever become dynamical.\footnote{Recall that in this section we take $M_\psi, M_S<\mu_0$ with $\mu_0$ from \eq{5TeV}.}

With $\psi$ and $S$ interacting, however, the putative Landau pole can be shifted beyond the Planck scale as illustrated in \fig{runBM1}, which shows the running couplings for a particular set of parameters. 
Specifically, we observe that some of the couplings settle into a slow walking regime by about $10^4$~TeV, in particular $\alpha_4$, but also the BSM Yukawa and  most quartics. The Higgs portal couplings $\alpha_{\delta,\dt}$ and the quartic Higgs coupling join the walking regime at ${\cal O}(10^5)$ TeV and ${\cal O}(10^6)$ TeV, respectively. Note, the portal couplings $\alpha_{\delta}$ and $\alpha_w$ change their sign during the RG evolution.
The SM gauge couplings continue to run moderately over the whole range of energies. Kinetic mixing and the top-Yukawa grow more strongly and, together with the BSM quartics, approach values around ${\cal O}(10^{-2}-10^{-1})$  at the Planck scale. An exception to this is the Higgs portal $\alpha_\delta$ that stays tiny as ${\cal O}(10^{-4})$.

Next, we perform a scan over roughly 75,000 different sets of initial conditions \eq{free} with $\psi$ and $S$ fully interacting, 
which establishes two  main parameter constraints.
First, we find a lower bound on the magnitude of the BSM Yukawa interaction
\begin{equation}
\label{eq:BM1-cond}
10^{-1.25} \lesssim \,  \alpha_y(\mu_0)\,,
\end{equation}
in accord with the estimate \eq{boundYukawa}. Second, we also find  that
\begin{equation}\label{eq:BM1-cond2}
\begin{aligned}10^{-4} \lesssim \alpha_{\delta}(\mu_0) \lesssim 10^{-0.75}\\{\rm or}\quad  \, 10^{-5} \lesssim \alpha_{\dt}(\mu_0) \lesssim 10^{-0.75}\,
\end{aligned}
\end{equation}
helps to promote Planck safety.
Note that it is sufficient if one of the Higgs portals fulfills one of the condition \eq{BM1-cond2} while the other one can be chosen almost freely.
As such, the range  \eq{BM1-cond2} is wider than the rough estimate \eq{range}.
The remaining BSM couplings are much more loosely constrained, as long as they do not interfere with the walking regime.

Our findings are further illustrated in \fig{ParameterSpaceBM1}, which shows the vacuum structure of models at the Planck scale, encoded in terms of the couplings 
$\alpha_y, \alpha_\delta$ and $\alpha_\dt$ at the matching scale. The color coding relates to whether poles have arisen (red),  whether the vacuum is unstable (gray)  or stable with vacuum $V^+$ (blue), or whether the Higgs is unstable ($\alpha_\lambda <-10^{-4}$, brown) or metastable ($-10^{-4}<\alpha_\lambda <0$, yellow).

By and large, we find that many low-energy parameters are excluded due to Landau poles, vacuum instability, or an enhanced instability in the Higgs sector. However, we also find a fair range of viable  settings with a stable ground state at the Planck scale. 
The corresponding Planck-safe trajectories are often similar to those in \fig{runBM1}, though  specific details such as the onset of walking may differ.
Interestingly though, we  find that only the most symmetric vacuum configuration $V^+$ is realized. For some  parameter ranges,  however, we observe intermediate transitions from the vacuum $V^+$ to the symmetry-broken vacuum  configuration $V^-$ along the trajectory. This transition is typically first order and could have left a trace in $e.g.$~cosmological data.

Finally, we recall that \BM1 does not require right-handed neutrinos, while the other benchmarks  do. 
For comparison, we have also studied a variant of \BM1 which instead uses  right-handed neutrinos  with all other specifics the same. While the putative Landau pole is  slightly lowered (to around 70~TeV), we find that the scan for Planck-safe trajectories gives a result quite similar to Figs.~\ref{fig:runBM1} and~\ref{fig:ParameterSpaceBM1}. We conclude that the formulation of \BM1 with or without  right-handed neutrinos is not instrumental for its UV behavior.

\subsection{Benchmark 2}

\begin{figure}[b]
  \centering
  \includegraphics[trim=1.5cm 0 1.5cm 0, clip, width=0.9\columnwidth]{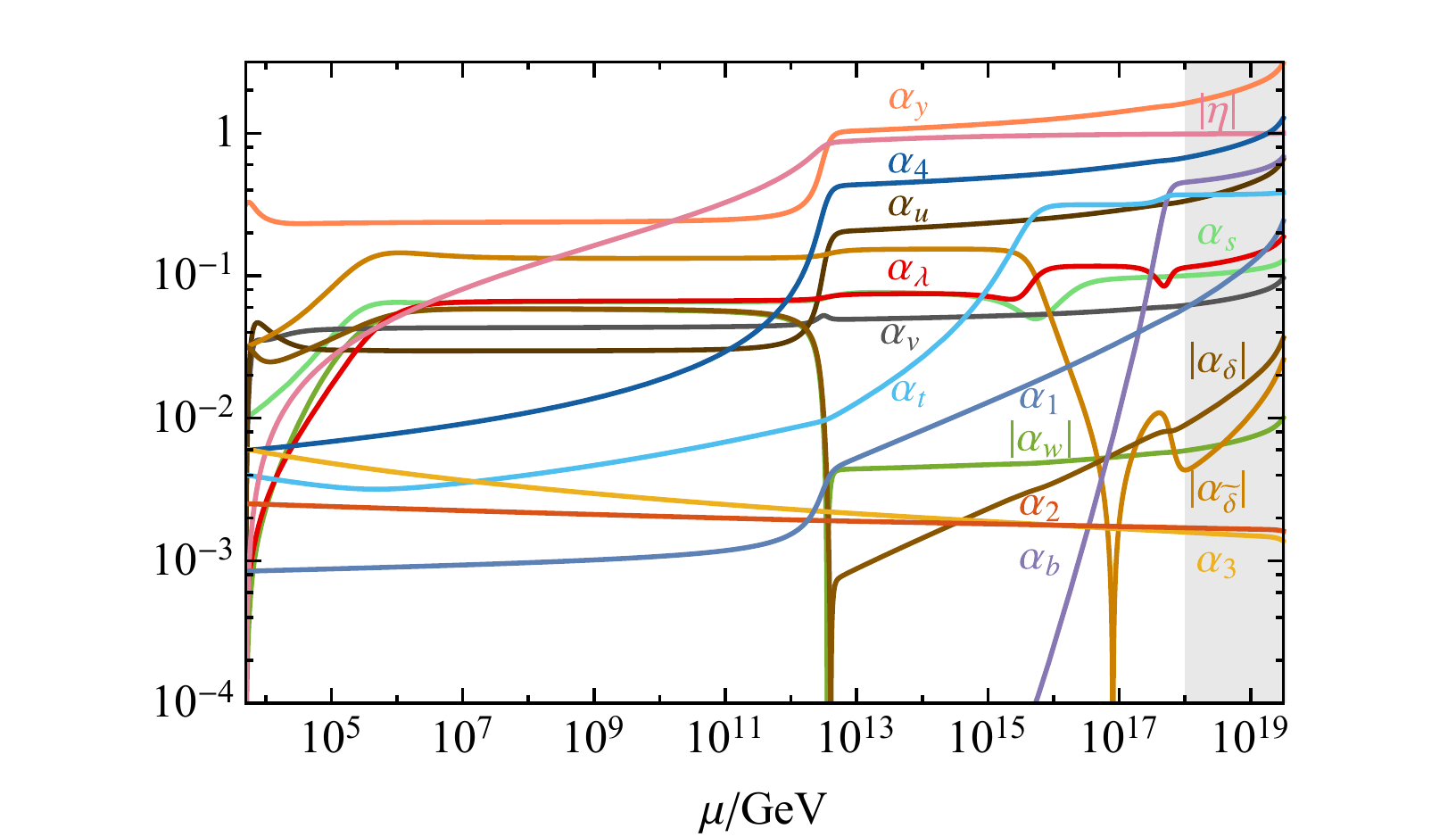}\\
  \caption{Sample running of couplings for \BM2 up to the Planck scale (gray area) showing trajectories for all couplings and kinetic mixing $\eta$
with initial conditions 
 \eq{5TeV}, \tab{SM-match}   alongside
   $\{\alpha_4,\eta, \alpha_{y,}, \alpha_{\delta},\, \alpha_{{\dt}},\, \alpha_u, \, \alpha_v, \, \alpha_w, \, \alpha_s \} |_{\mu_0} = \{5.97 \cdot 10^{-3},0, 10^{-0.5}, 10^{-1.5},10^{-1.5}, 10^{-4},  10^{-5},  10^{-6} ,  10^{-2}\}$.}
  \label{fig:runBM2}
\end{figure}

\begin{figure}[h!]
  \centering
  \renewcommand*{\arraystretch}{0}
  \begin{tabular}{c}
    \includegraphics[width=.9\columnwidth]{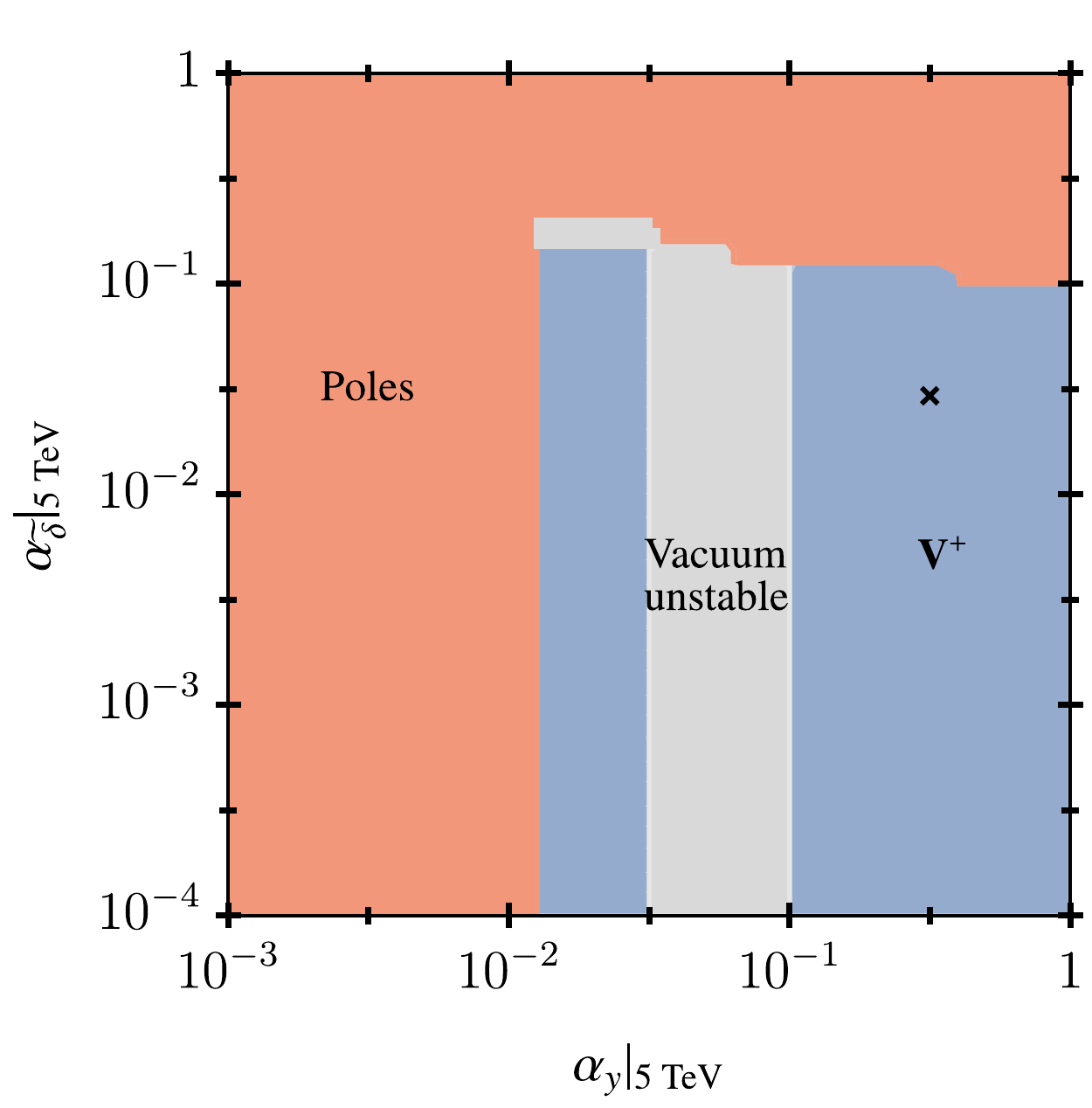}\\
    \includegraphics[width=.9\columnwidth]{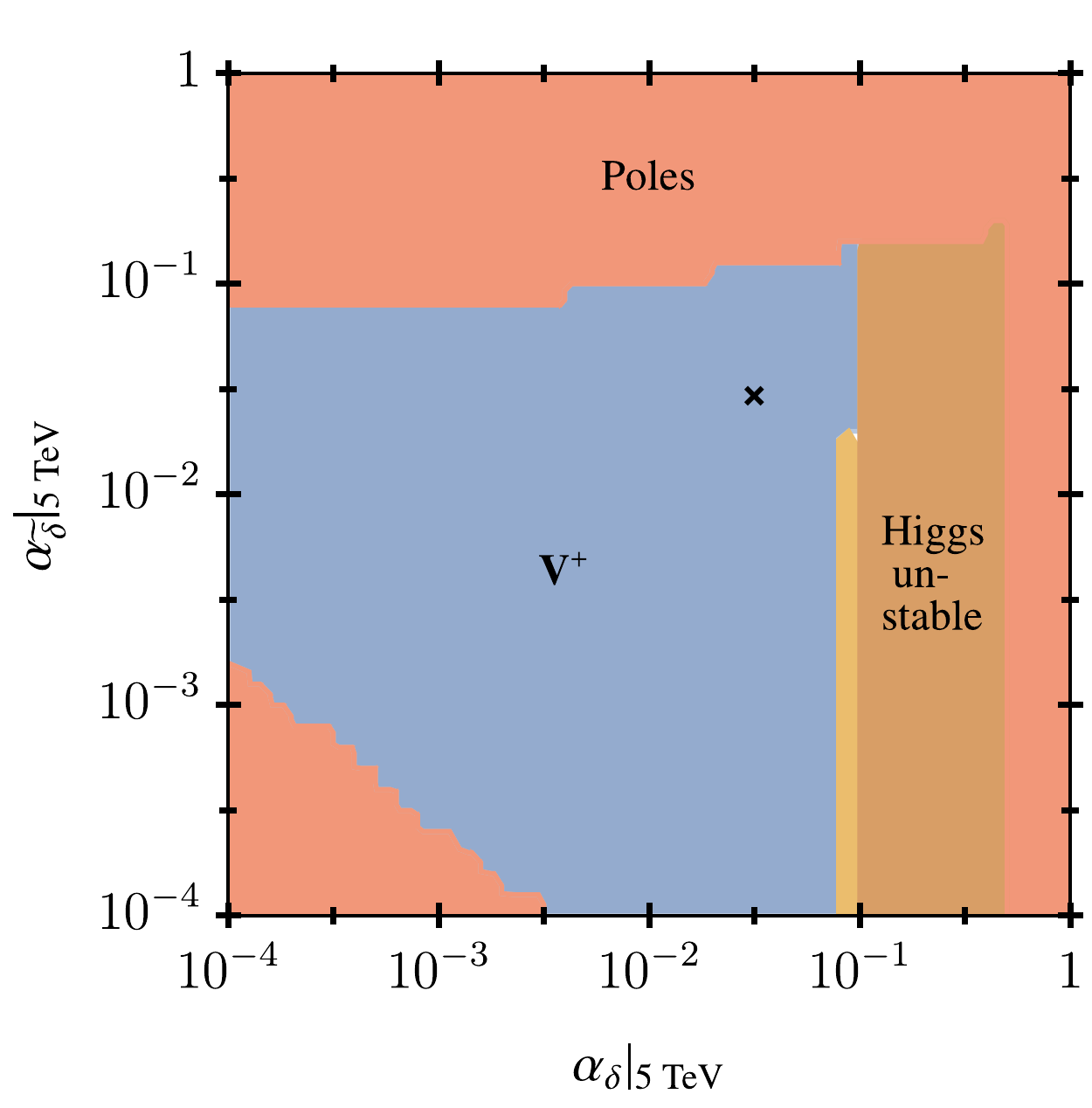}\\
  \end{tabular}
  \caption{Critical surface of parameters for the benchmark model \BM2, projected onto the $\{\alpha_{y},\alpha_{\dt}\}|_{\mu_0}$  (top) and the $\{\alpha_{\delta},\alpha_{\dt}\}|_{\mu_0}$ plane of parameters (bottom). We further use  the matching scale \eq{5TeV} and \tab{SM-match},
  $\alpha_\delta|_{\mu_0}=10^{-1.5}$ (top), $\alpha_y|_{\mu_0}=10^{-0.5}$ (bottom), together with $\{\,\alpha_4,\, \eta, \,\alpha_u, \, \alpha_v, \, \alpha_w, \, \alpha_s \} |_{\mu_0}= \{5.97 \cdot 10^{-3}, 0, 10^{-4},  10^{-5},  10^{-6},  10^{-2}\}$.
  The black cross indicates the sample trajectory in \fig{runBM2}; colors as in \fig{ParameterSpaceBM1}.}
  \label{fig:ParameterSpaceBM2}
\end{figure}

In the benchmark model \BM2 the  $\alpha_4$-value required   to explain the $B$-anomalies reads
$\alpha_4(\mu_0) = 5.97 \cdot 10^{-3}$,
which implies a na\"ive estimate for a Landau pole at around $2\cdot10^5$~TeV. Again, this scale 
also provides a theoretical upper bound for the  mass of the $\psi$ and $S$ fields. 

Further, \fig{runBM2} illustrates that a Landau pole can be avoided. For the chosen  set of parameters we observe that couplings initially  settle within a walking regime while subsequently crossing over into a secondary walking regime around $10^9$ TeV, and ultimately settling in the vacuum state $V^+$. 
SM gauge couplings continue to run moderately and the Higgs remains stable throughout, while some of the portal couplings even change sign. Similarly to \fig{runBM1}, we observe that some of the couplings
approach values of order unity around Planckian energies while others remain much smaller.

More generally, scanning over a vast set of initial conditions \eq{free}, we find the  constraints 
\begin{equation}
\label{eq:BM2-cond}
10^{-1.75} \lesssim \,  \alpha_y(\mu_0)\,,
\end{equation}
in accord with the estimate \eq{boundYukawa}, and together with
\begin{equation}\label{eq:BM2-cond2}
\begin{aligned}&10^{-6} \lesssim \alpha_{\delta}(\mu_0) \lesssim 10^{-1}\\{\rm or}  \quad &10^{-5} \lesssim \alpha_{\dt}(\mu_0) \lesssim 10^{-0.5}\,.
\end{aligned}
\end{equation}
Once more, the range  \eq{BM2-cond2} comes out larger than the estimate \eq{range}, and
only one of the  Higgs portals has to fulfill one of the condition \eq{BM2-cond2}. 
The remaining BSM couplings are less constrained, as long as they do not interfere with the walking regime.

Results are further illustrated in \fig{ParameterSpaceBM2}, which shows the vacuum structure of models at the Planck scale, encoded in terms of the couplings $\alpha_y, \alpha_\delta$ and $\alpha_\dt$ at the matching scale, also using the same  color coding as in \fig{ParameterSpaceBM1}.
On the whole, we find that many settings are excluded due to  poles and instabilities.
Still, we again find a fair range of viable settings with a stable ground state $V^+$ at the Planck scale. 
Planck-safe trajectories look similar to those in \fig{runBM2}, though some of the specifics including the onset  of walking  may differ. 

Finally, while many couplings remain small, we   note that a few of them, in particular the BSM Yukawa, can become  large if not outright non-perturbative.
This is not entirely unexpected, and a response to the fact that the $B$-anomalies  necessitate relatively large $\alpha_4$ already at the matching scale, see \eq{g4bound}.  Then, in order to delay the Landau pole dynamically,  this evidently triggers   compensating interactions of similar strength.
For this reason, some of our results must be taken with a grain of salt as higher order corrections may  well affect the critical surface  quantitatively. 
Still, we find a coherent picture overall, where the requisite Yukawas only differ by a factor of a few between  benchmark models.

\subsection{Benchmark 3}

\begin{figure}[t]
  \centering
  \renewcommand*{\arraystretch}{0}
  \begin{tabular}{c}
    \includegraphics[width=.9\columnwidth]{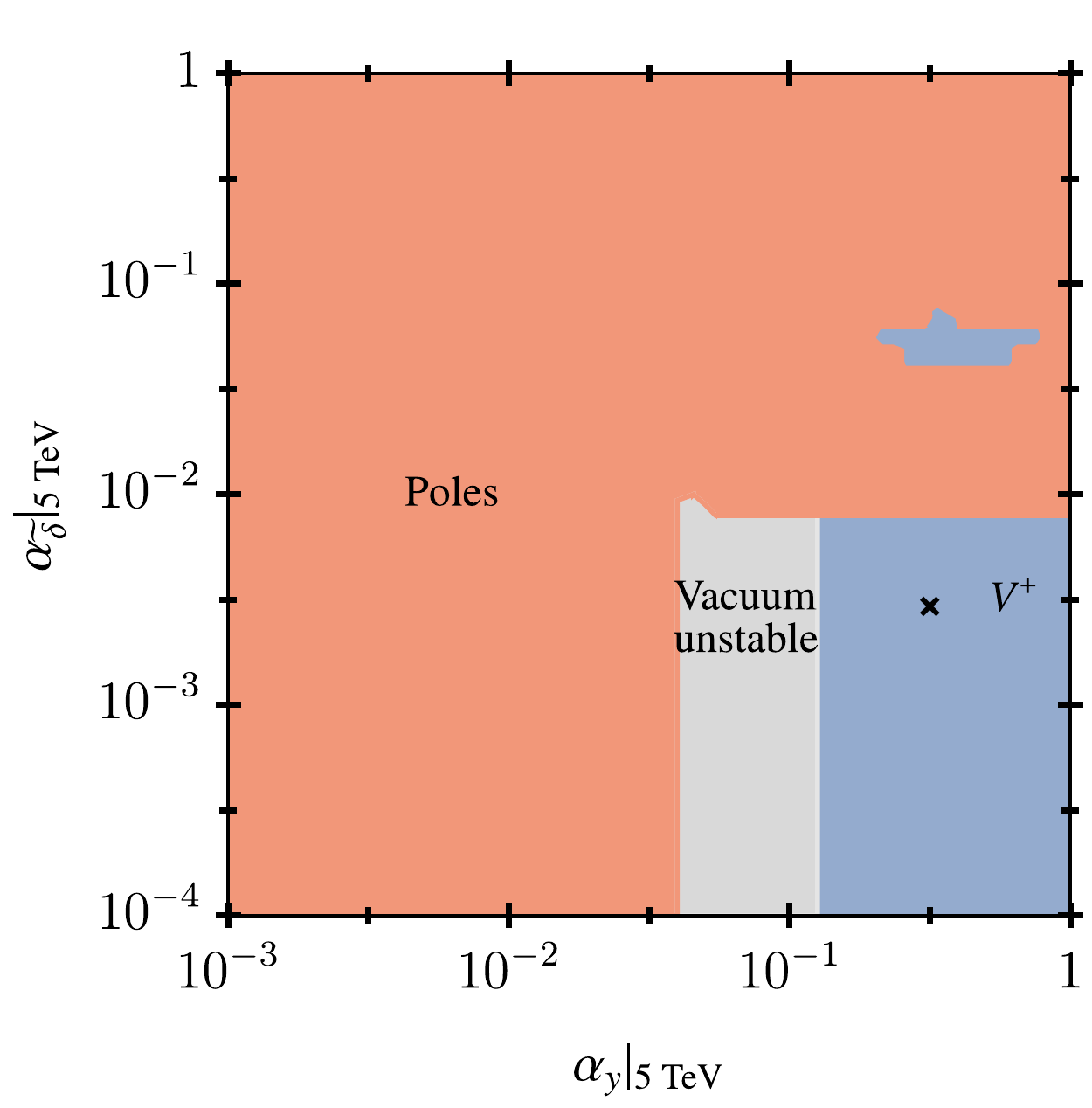}\\
    \includegraphics[width=.9\columnwidth]{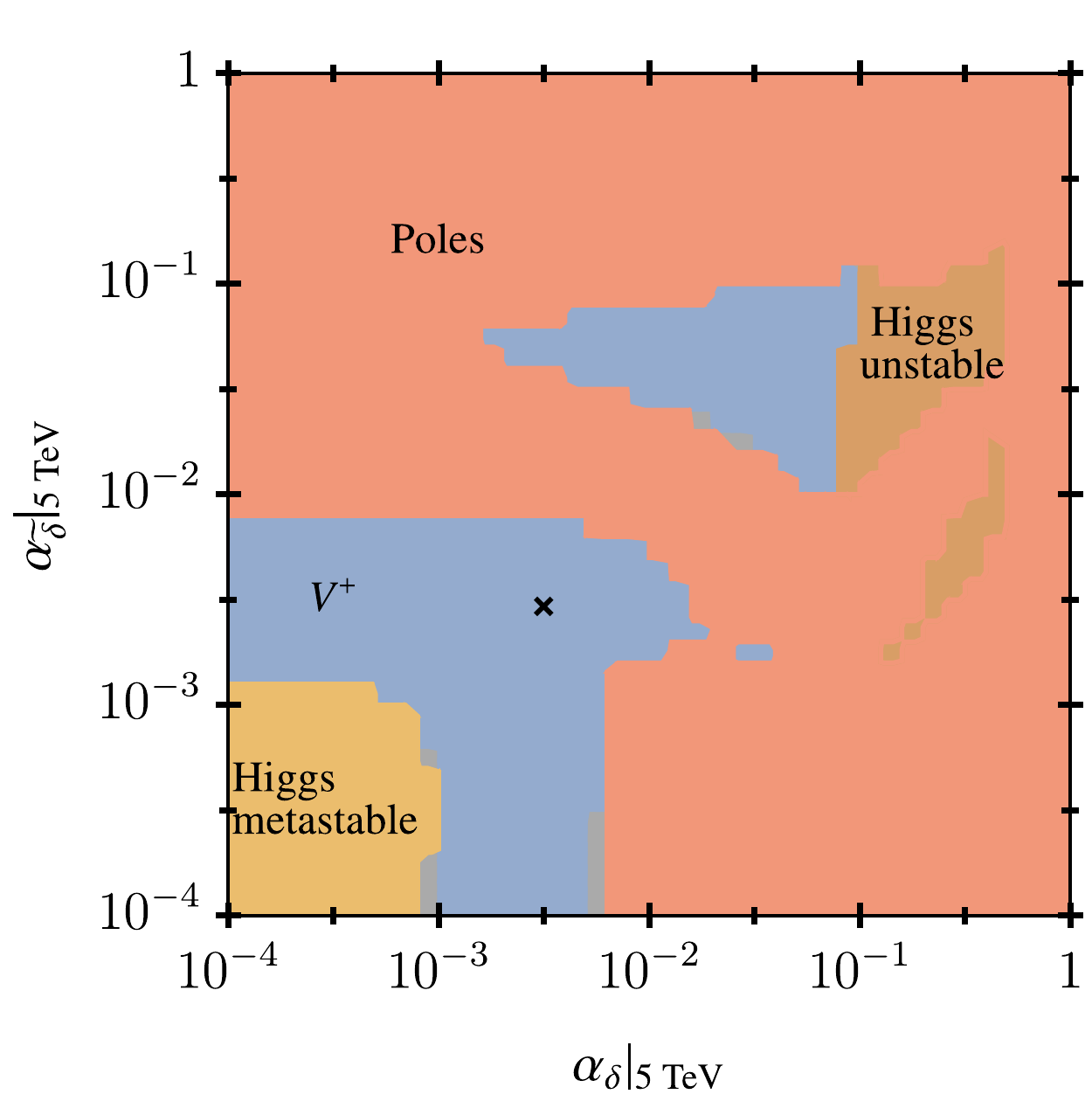}\\
  \end{tabular}
  \caption{Critical surface of parameters for  the benchmark model \BM3, projected onto the $\{\alpha_{y},\alpha_{\dt}\}|_{\mu_0}$  (top) and the $\{\alpha_{\delta},\alpha_{\dt}\}|_{\mu_0}$ plane of parameters (bottom). We further use  the matching scale \eq{5TeV} and \tab{SM-match},
  $\alpha_\delta|_{\mu_0}=10^{-2.5}$ (top), $\alpha_y|_{\mu_0}=10^{-0.5}$ (bottom), together with $\{\,\alpha_4,\, \eta, \,\alpha_u, \, \alpha_v, \, \alpha_w, \, \alpha_s \} |_{\mu_0}= \{4.60\cdot 10^{-2}, 0,  10^{-4},  10^{-5} ,  10^{-6},  10^{-3.5}\}$. 
  The black cross indicates the sample trajectory in \fig{runBM3}; colors as in \fig{ParameterSpaceBM1}.}
  \label{fig:ParameterSpaceBM3}
\end{figure}

\begin{figure}
  \centering
  \includegraphics[trim=1.5cm 0 1.5cm 0, clip, width=0.9\columnwidth]{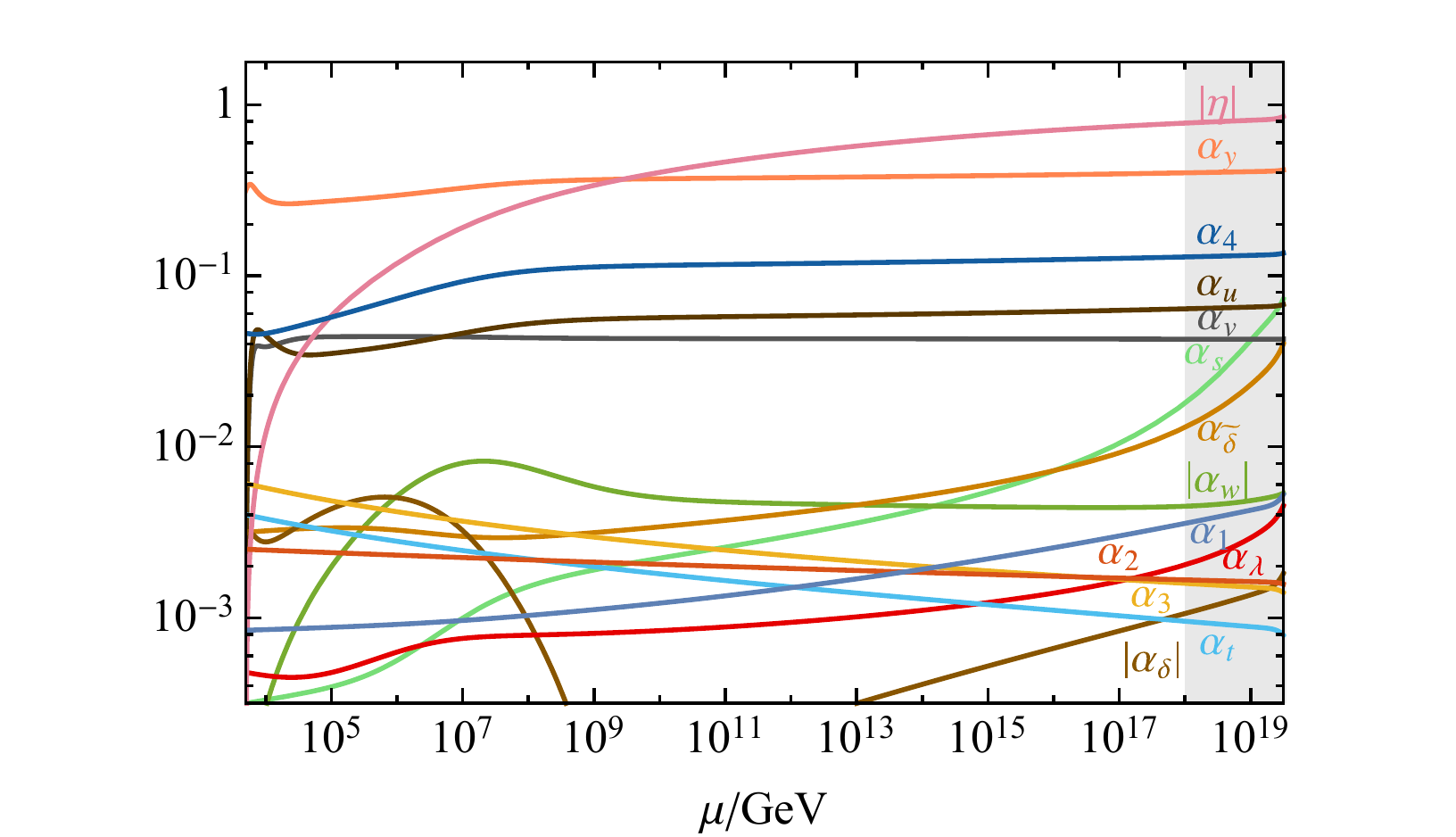}\\
  \caption{Sample  running of couplings for \BM3 up to the Planck scale (gray area) showing trajectories for all couplings and kinetic mixing $\eta$
with initial conditions  \eq{5TeV}, \tab{SM-match}   alongside
$\{\alpha_4,\eta, \alpha_{y,}, \alpha_{\delta},\, \alpha_{{\dt}},\, \alpha_u, \, \alpha_v, \, \alpha_w, \, \alpha_s \} |_{\mu_0} = \{4.60 \cdot 10^{-2},0, 10^{-0.5}, 10^{-2.5},10^{-2.5}, 10^{-4},  10^{-5},  10^{-6} ,  10^{-3.5}\}$.}
  \label{fig:runBM3}
\end{figure}

For benchmark \BM3, the  $\alpha_4$-value required  for the $B$-anomalies
$\alpha_4(\mu_0) = 4.60 \cdot 10^{-2}$ implies a  putative Landau pole around $25$~TeV,
which coincides with the theoretical upper bound  for the  mass of the $\psi$ and $S$ fields in \BM3.
Scanning as before over a large set of initial conditions \eq{free}, we find the general constraint 
\begin{equation}
\label{eq:BM3-cond}
10^{-1} \lesssim \,  \alpha_y(\mu_0)\,,
\end{equation}
in accord with the estimate \eq{boundYukawa}, and together with
\begin{equation}\label{eq:BM3-cond2}
\begin{aligned}&10^{-4}\ \ \lesssim \alpha_{\delta}(\mu_0) \lesssim 10^{-1}\\{\rm or}  \quad &10^{-2.5} \lesssim \alpha_{\dt}(\mu_0) \lesssim 10^{-1}\,.
\end{aligned}
\end{equation}
As expected, the range  \eq{BM3-cond2} comes out larger than the estimate \eq{range}, and
 it is sufficient, once more, if one of the Higgs portals 
fulfills one of the conditions \eq{BM3-cond2} while the other one can be chosen  freely, with the remaining 
BSM couplings largely unconstrained.

Our results are further illustrated in \fig{ParameterSpaceBM3}, which shows the vacuum of models at the Planck scale, encoded in terms of the couplings $\alpha_y, \alpha_\delta$ and $\alpha_\dt$ at the matching scale and using the same  color coding as in \fig{ParameterSpaceBM1}.
While many settings are excluded due to  poles and instabilities, we again find a fair range of settings where a stable ground state $V^+$ prevails.
What's new compared to \BM1 and \BM2 is that the UV critical surface of parameters is cut into two disconnected pieces, separated by the occurrence of poles for the parameters in between, which is visible in both projections.
Also, both Higgs portal couplings become sizable at the matching scale in one of the regions, but still within the bounds \eq{BM3-cond2}, while the BSM Yukawa  is sizable in both of them, see \eq{BM3-cond}.

\fig{runBM3} shows a sample trajectory   of \BM3, with parameters corresponding to the cross in  \fig{ParameterSpaceBM3}. 
 On the whole, all couplings evolve very slowly, but still noticeably, between the matching and the Planck scale. 
 In comparison with  the sample trajectories for \BM1 and \BM2  (see \fig{runBM1} and  \fig{runBM2}, respectively), we note that  the walking regime  is  somewhat less pronounced.
 At Planckian energies,  the kinetic mixing becomes of order unity, and the $U(1)'$ coupling, and the BSM Yukawa and quartics,  reach values of order ${\cal O}(10^{-1})$. All other couplings remain roughly within the range ${\cal O}(10^{-3}-10^{-2})$. Similar trajectories are found in  other Planck-safe parameter regions of \fig{ParameterSpaceBM3}.

\subsection{Benchmark 4}

\begin{figure}[t]
  \centering
  \renewcommand*{\arraystretch}{0}
  \begin{tabular}{c}
    \includegraphics[width=.9\columnwidth]{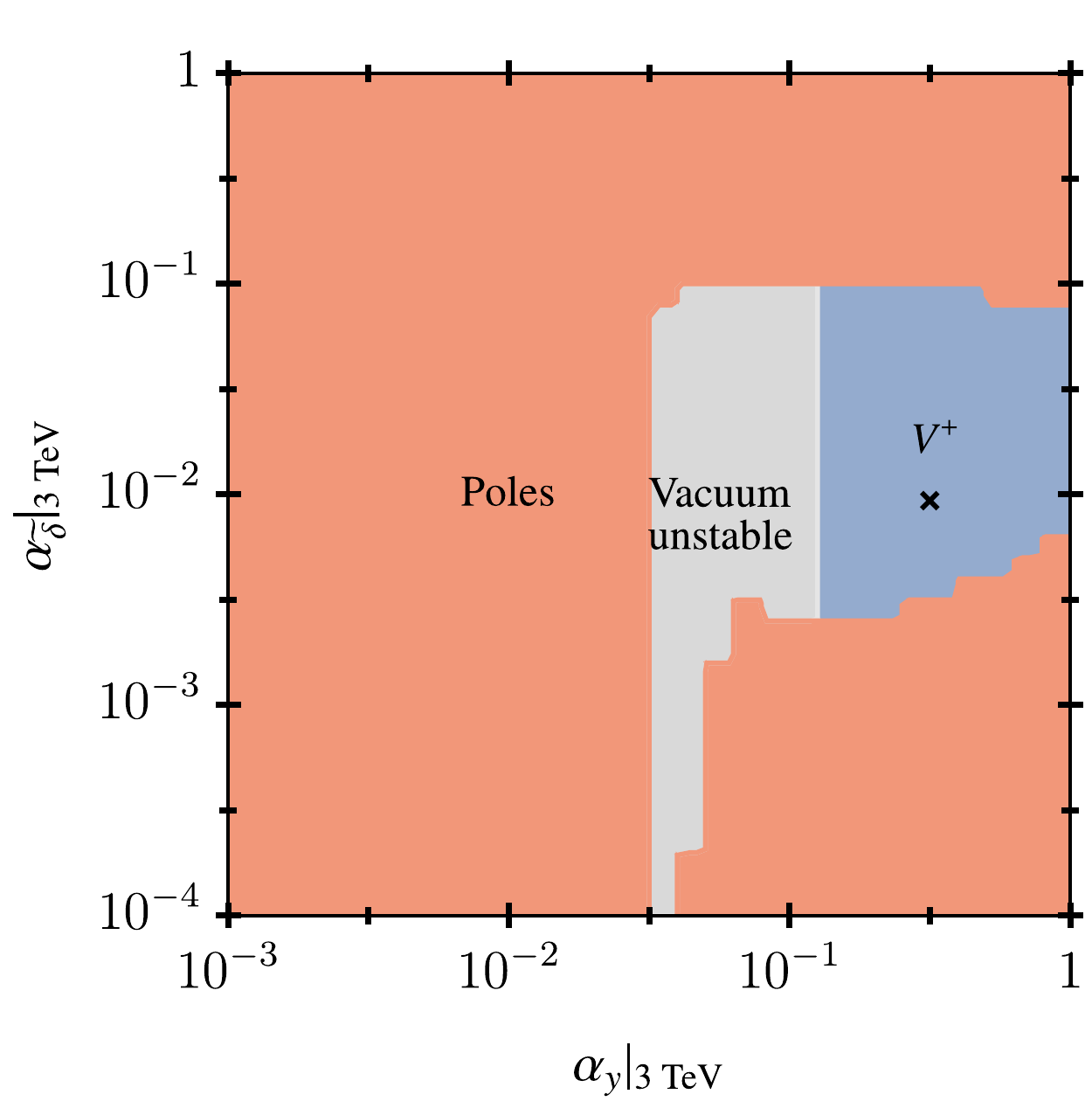}\\
    \includegraphics[width=.9\columnwidth]{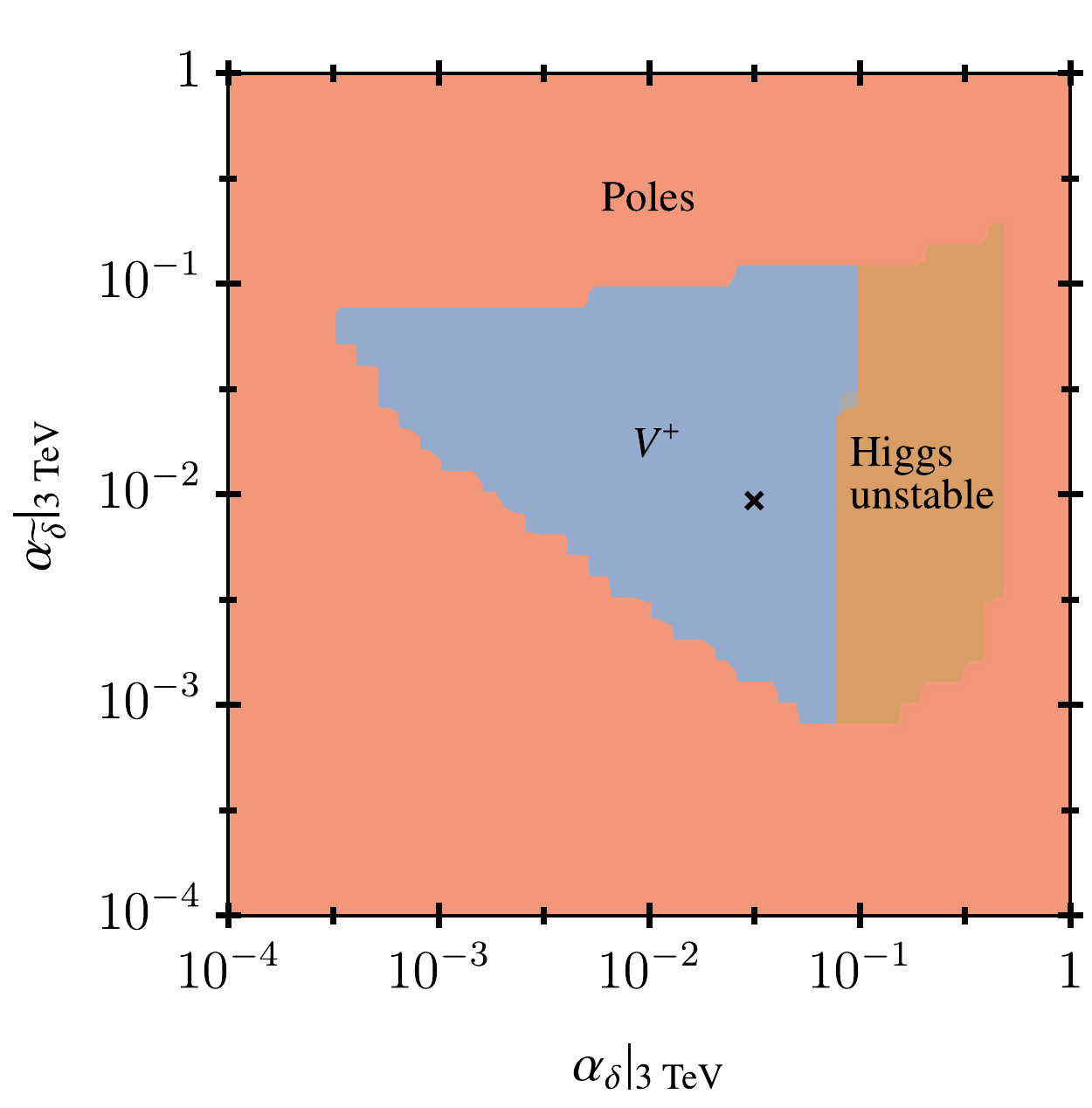}\\
  \end{tabular}
  \caption{Critical surface of parameters for  the benchmark model \BM4, projected onto the $\{\alpha_{y},\alpha_{\dt}\}|_{\mu_0}$  (top) and the $\{\alpha_{\delta},\alpha_{\dt}\}|_{\mu_0}$ plane of parameters (bottom). We further use the matching scale \eq{5TeV} and \tab{SM-match},
  $\alpha_\delta|_{\mu_0}=10^{-1.5}$ (top), $\alpha_y|_{\mu_0}=10^{-0.5}$ (bottom), together with $\{\,\alpha_4,\, \eta, \,\alpha_u, \, \alpha_v, \, \alpha_w, \, \alpha_s \} |_{\mu_0}= \{2.45\cdot 10^{-2}, 0,  10^{-4},  10^{-5} ,  10^{-6},  10^{-2}\}$. 
  The black cross indicates the sample trajectory in \fig{runBM4}; colors as in \fig{ParameterSpaceBM1}.}
  \label{fig:ParameterSpaceBM4}
\end{figure}

\begin{figure}
  \centering
  \includegraphics[trim=1.5cm 0 1.5cm 0, clip, width=0.9\columnwidth]{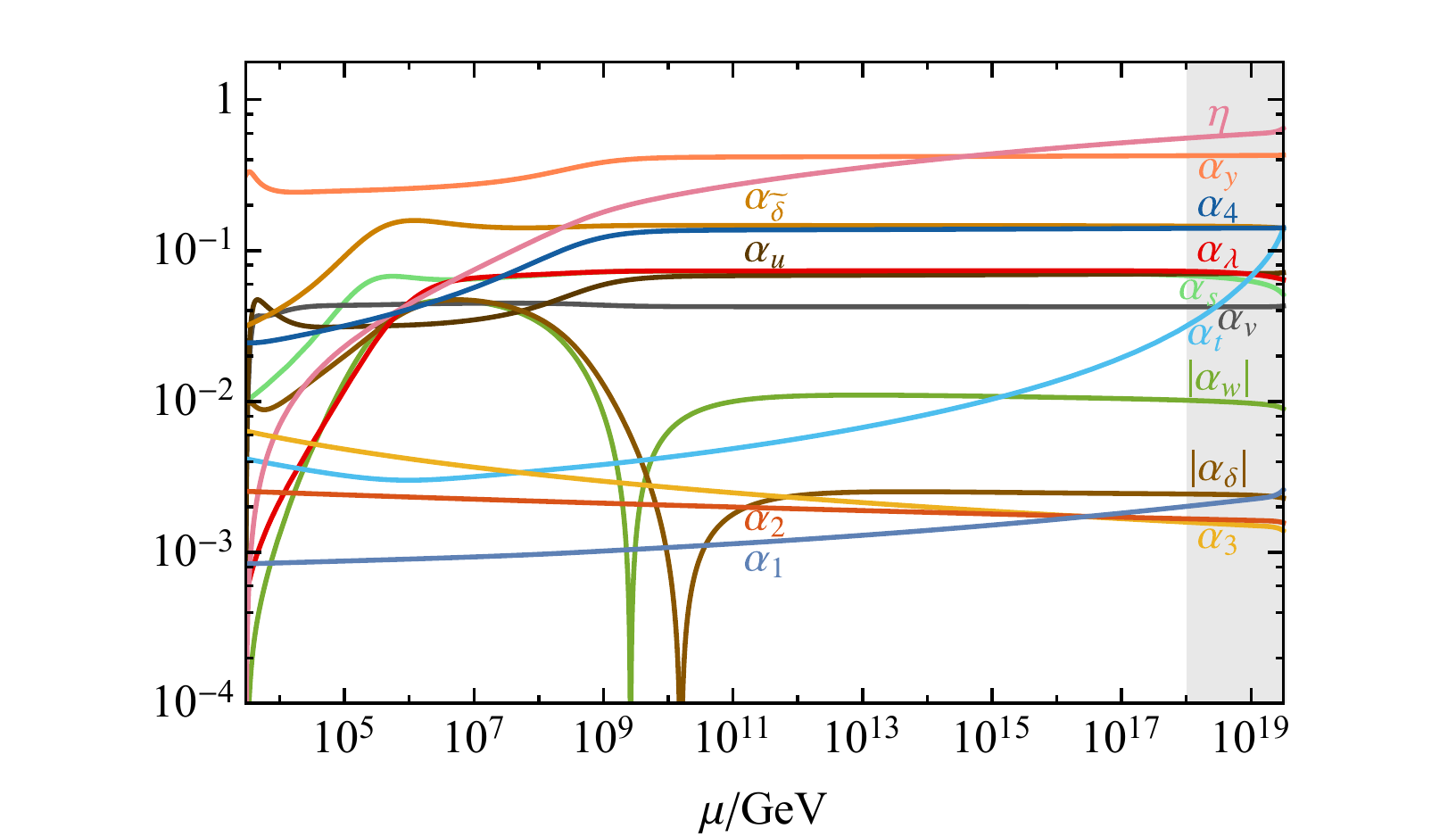}\\
  \caption{Sample  running of couplings for \BM4 up to the Planck scale (gray area) showing trajectories for all couplings and kinetic mixing $\eta$
with initial conditions \eq{5TeV}, \tab{SM-match} alongside
$\{\alpha_4,\eta, \alpha_{y,}, \alpha_{\delta},\, \alpha_{{\dt}},\, \alpha_u, \, \alpha_v, \, \alpha_w, \, \alpha_s \} |_{\mu_0} = \{2.45 \cdot 10^{-2},0, 10^{-0.5}, 10^{-1.5},10^{-2}, 10^{-4},  10^{-5},  10^{-6} ,  10^{-2}\}$.}
  \label{fig:runBM4}
\end{figure}

For benchmark \BM4 we have fixed the matching scale as $\mu_0 = 3$ TeV. At this scale, $\alpha_4(\mu_0) = 2.45 \cdot 10^{-2}$ accounts for the $B$-anomalies and  implies a  putative Landau pole around $60$~TeV, yielding again a theoretical upper bound  for the  mass of the $\psi$ and $S$ fields in \BM4.
 Scanning as before over a large set of initial conditions \eq{free}, we find the general constraints 
\begin{equation}
\label{eq:BM4-cond}
10^{-1.25} \lesssim \,  \alpha_y(\mu_0)\,,
\end{equation}
in accord with the estimate \eq{boundYukawa}. Moreover, Planck safety is promoted by at least one of the Higgs portal couplings fulfilling
\begin{equation}\label{eq:BM4-cond2}
\begin{aligned}&10^{-4} \lesssim \alpha_{\delta}(\mu_0) \lesssim 10^{-1}\\{\rm or}  \quad &10^{-4} \lesssim \alpha_{\dt}(\mu_0) \lesssim 10^{-1}\,.
\end{aligned}
\end{equation}
Additionally, 
we also find Planck-safe trajectories with both $|\alpha_{\delta,\dt}(\mu_0)|$ being tiny or even zero. This is possible as $\alpha_\dt$ is switched on radiatively by a contribution to its $\beta$-function $\propto \alpha_4^2 \alpha_t$ due to the non-vanishing $U(1)'$ charge of $\phi$. $\alpha_\dt$ then also radiatively induces $\alpha_\delta$. If this happens such that $\alpha_{\delta,\dt}$ become sizable quickly enough, the vacuum can be stabilized although $|\alpha_{\delta,\dt}(\mu_0)|$ were tiny. 

Our results are further illustrated in \fig{ParameterSpaceBM4}, which shows the vacuum configuration at the Planck scale  in terms of the couplings $\alpha_y, \alpha_\delta$ and $\alpha_\dt$ at the matching scale and using the same  color coding as in \fig{ParameterSpaceBM1}.
While many settings are excluded due to  poles and instabilities, we again find a fair chunk of parameter space where a stable ground state $V^+$ prevails. The BSM critical surface in the plane of $\alpha_y$ and $\alpha_\dt$ appears similar to \BM1, while in the plane of $\alpha_\delta$ and $\alpha_\dt$ the BSM critical surface has qualitatively the same shape as in \BM3, although the allowed window for $\alpha_{\delta,\dt}$ is a bit more narrow.

\fig{runBM4} shows a sample trajectory of \BM4, with parameters corresponding to the cross in  \fig{ParameterSpaceBM4}. Most couplings enter a walking regime at $\mathcal{O}(10^6)$ GeV while the portals $\alpha_{\delta, w}$ get negative and join the walking regime at higher energies of $\mathcal{O}(10^{11})$ GeV.

At Planckian energies, the kinetic mixing $\eta$ and the BSM Yukawa $\alpha_y$ become of order unity. The $U(1)'$ coupling $\alpha_4$ as well as $\alpha_t$ and most quartics reach values of order ${\cal O}(10^{-1})$. In contrast, the portal couplings $\alpha_{\delta,w }$ as well as the SM gauge couplings arrive at moderate values of ${\cal O}(10^{-3})$, whereas $\alpha_b$ stays tiny all the way up to the Planck scale and therefore does not appear in the plotted range of \fig{runBM4}.

\section{\bf  Phenomenological Implications}
\label{sec:Discussion}

In the previous sections we  presented phenomenologically viable benchmark models,
which account for the $B$-anomalies and remain predictive up to the  Planck scale in suitable parameter regions. Here, we discuss phenomenological implications, starting with predictions for dineutrino modes (Sec.~\ref{sec:nunu}), followed by collider signatures (Sec.~\ref{sec:collider}).
We also entertain the possibility of right-handed Wilson coefficients (Sec.~\ref{sec:RH-fcnc}).

\subsection{Predictions for \texorpdfstring{$B \to K^{(*)} \nu \bar \nu$}{B->K(*) nu nu} \label{sec:nunu}}

With the charge assignments as in \tab{Benchmarks} and Wilson coefficients  as in \Sec{Benchmarks},
predictions for dineutrino branching ratios can be obtained.
Since right-handed quark currents are not necessary and neglected, the impact of the $Z^\prime$ models is universal for all $B \to H \nu \bar \nu$, $H=K,K^*,\ldots$ branching ratios \cite{Bause:2021ply}
\begin{align} \label{eq:BHvv_ratio}
&\frac{ \mathcal{B}\left(B \to H \nu \bar{\nu}\right)}{ \mathcal{B}\left(B \to H \nu \bar{\nu}\right)_\text{SM}} =\frac{1}{3}\left(\,\sum_\ell |1+F_{L_\ell} a |^2 + |F_{\nu_\ell} a |^2\right)\,,\\ 
a &= 2\,\mathcal{N}^{-1} g_L^{bs} g_4/(M_{Z^\prime}^2 \, X_\text{SM})\,, 
\quad |a|\ll 1 \,, \nonumber
\end{align}
and reads $1.003$, $1.05$, $1.08$ and $0.97$ for \BM{1}, \BM{2}, \BM{3} and \BM{4}, respectively.
Above, $X_\text{SM}=-12.64$ \cite{Brod:2021hsj} denotes the corresponding Wilson coefficient in the SM.
The smallness in \BM{1} stems from a cancellation between the second and third generations with $F_{L_2}=-F_{L_3}$ in the interference terms with the SM.
Contributions from light right-handed neutrinos have no interference with the SM, see also \cite{Buras:2014fpa}, and their impact on \eq{BHvv_ratio} is at the permille level, and negligible.
We observe, as expected \cite{Bause:2021ply}, that the $B$-anomalies generically support an enhancement of the $b \to s$ dineutrino branching ratios although a very mild suppression is obtained in \BM{4} where $a>0$ and $F_{L_3}<0$.
The Belle II experiment is expected to observe $B \to K \nu \bar{\nu}$ and $B \to K^\ast \nu \bar{\nu}$ at the SM-level \cite{Belle-II:2018jsg},
hence in all benchmarks. 
Study of dineutrino
branching ratios becomes then informative to distinguish
concrete models. We note that the NP effects in our benchmarks are too small to be distinguished from the predictions of the SM within present precision.

\subsection{Collider Signatures \label{sec:collider}}

The $Z^\prime$ boson  can decay to fermion-antifermion pairs. 
The corresponding decay width reads \cite{Kang:2004bz}
\begin{equation}\label{eq:ZPrff}
\begin{aligned}
&\Gamma(Z^\prime \to f_i\bar{f}_i)
=  \frac{2\pi N_C^f}{3} \alpha_4 M_{Z^\prime} \sqrt{1-4\frac{m_{f_i}^2}{M_{Z^\prime}^2}} \\ & \cdot \left[ (F_{f_{Li}}^2 + F_{f_{Ri}}^2) - \frac{m_f^2}{M_{Z^\prime}^2}  (F_{f_{Li}}^2 -6 F_{f_{Li}} F_{f_{Ri}} + F_{f_{Ri}}^2) \right] 
\end{aligned}
\end{equation}
where  kinetic mixing has been neglected, and with color factor
$ N_C^f=3$ for quarks and  $N_C^f=1$ otherwise. 
If $M_{Z^\prime}>2 M_\psi$, the decay to $\psi \bar{\psi}$ is kinematically allowed and becomes dominant. 
Note, phase space suppression of the  partial decay width $\Gamma(Z^\prime \to \psi\bar{\psi})$  only becomes relevant ($>5\%$) for $M_\psi \gtrsim 0.3 \, M_{Z^\prime}$.

When neglecting kinetic mixing, the $Z^\prime$ cannot decay to SM gauge bosons at leading order.  However, the decay $Z^\prime \to ss$ with $s=h,\phi$ is possible if $F_{s} \neq 0$. The decay width is given by~\cite{Kang:2004bz} 
\begin{equation}\label{eq:ZPrSS}
\Gamma(Z^\prime \to ss) \simeq  \frac{\pi}{3} \alpha_4 M_{Z^\prime}  F_s^2\left( 1 - 4\frac{m_s^2}{M_{Z^\prime}^2} \right)^{3/2}.
\end{equation}

Summing  the partial decay widths in the limit $M_\psi \lesssim 0.3 \, M_{Z^\prime}$ yields a total $Z^\prime$-width of
\begin{align}
    \Gamma^\text{tot} (Z^\prime)\big|_{\text{\BM 1}}&= 0.43\, M_{Z^\prime}\,, \\
    \Gamma^\text{tot} (Z^\prime)\big|_{\text{\BM 2}}&= 0.14\, M_{Z^\prime}\,, \\
    \Gamma^\text{tot} (Z^\prime)\big|_{\text{\BM 3}}&= 0.73\, M_{Z^\prime}\,, \\
    \Gamma^\text{tot} (Z^\prime)\big|_{\text{\BM 4}}&= 0.43\, M_{Z^\prime}\,,
\end{align}
in the $Z^\prime$ benchmark models.
We learn that the $Z^\prime$ is generically quite broad, due to the sizable $U(1)^\prime$ coupling.
Note that the width is roughly halved (reduced by a factor of 4-5) in \BM {1,2} (\BM{3,4}) if $Z^\prime \to \psi \bar{\psi}$ is kinematically forbidden.

\begin{table}
\setlength\arrayrulewidth{1.3pt}
\def\arraystretch{1.1}
 \setlength{\tabcolsep}{1pt}
\centering
\begin{tabular}{|c|c|c|c|c|c|c|c|c|c|c|}
\hline
Model & jets & $b$ & $t$ & $e$ & $\mu$ & $\tau$ &  $\nu_{e,\mu,\tau}$ & $h$ & $\psi_{1,2,3}$ & $\phi$ \\ \hline
\BM 1 & 0.5 & 0.5 & 0.5 & 0 & 15 & 15 & 15 & 0 & 54 & 0.2 \\ 
\BM 2 & 14 & 1.5 & 1.5 & 0 & 9 & 9 & 18 & 0 & 46 & 0.1 \\ 
\BM 3 & 5 & 0 & 0 & 0 & 4 & 4 & 8 & 0 & 79 & 0.1 \\
\BM 4 & 0 & 0.9 & 0.9 & 0 & 3 & 11 & 14 & 0 & 72 & 0.2 \\\hline 
\end{tabular}
\caption{Tree-level branching fractions in \% for the different $Z^\prime$ decay modes to fermion-antifermion pairs and pairs of scalars neglecting fermionic as well as kinetic mixing. 
The numerical values correspond to the scenario where the decays $Z^\prime \to \psi_i \bar{\psi}_i, \phi \phi$ are kinematically allowed and hardly phase space suppressed (i.e. $M_{\psi,\phi} \lesssim 0.3 \, M_{Z^\prime}$). 
If the decay to $\psi\bar{\psi}$ is kinematically significantly suppressed or forbidden, the other branching rations increase by up to roughly a factor of 2, 4 and 5 in \BM {1,2}, \BM 3 and \BM4, respectively.
}
\label{tab:Brs}
\end{table}

The branching fractions $\mathcal{B}(Z^\prime \to f_i\bar{f}_i, ss)$ can be calculated from \eq{ZPrff}, \eq{ZPrSS} and are given in \tab{Brs}. 
If kinematically allowed, the decay mode $Z^\prime \to \psi \bar{\psi}$, where we summed over three generations, is generically dominant and accounts for a branching fraction of $\sim50\,\% \,\sim 80\,\%$ and $\sim 70\,\%$ in \BM {1,2}, \BM{3} and \BM4, respectively. 
Another common feature is a sizable branching fraction to dineutrinos, ranging from $8$ to $18\%$ in the concrete benchmark models. 
This includes both contributions from the upper (neutrino) component of $L$, as well as $\nu$.
Both decays $ Z^\prime \to \psi \bar{\psi},\nu \bar{\nu}$ yield the same signature, namely an invisible final state, i.e. missing energy, that can be searched for at the LHC. 
The strongly dominating decay mode $Z^\prime \to \text{invisible}$ with a branching ratio of $\sim 65 -  85 \%$ is the most outstanding feature of our $Z^\prime$ model class. 

The benchmarks feature different branching fractions to dimuons (and equally to ditaus) of $15\%$, $9\%$, $4\%$ and $3\%$ in \BM1, \BM2, \BM3 and \BM4, respectively. In \BM{1-3} decays to ditaus have the same branching ratio as to dimuons, whereas in \BM4 tauonic decays with a branching ratio of $11\, \%$ are four times more abundant than muonic ones. Decays to dielectrons are switched-off \eq{ElectronBound}.
Similarly, the models can be also distinguished from their branching ratio to dijets of $0\% - 14\%$. 
Hence, measuring $Z^\prime$ branching fractions with an accuracy of $\sim 10\%$ allows for a distinction between the benchmarks.

On the other hand, the decays  $Z^\prime \to b\bar{b}, t\bar{t}, hh, \phi \phi$ have branching fractions of $\lesssim 1\%$ in all models and are therefore negligible.

The $Z^\prime$ Drell-Yan cross section at a hadron machine can be approximated as \cite{Paz:2017tkr,Zyla:2020zbs}
\begin{equation}
    \sigma(pp\to Z^\prime X \to f\bar{f} X) \simeq \frac{\pi}{6 s} \sum_q c_q^f w_q(s,{M_{Z^\prime}}^2)
\end{equation}
where $q=u,d,s,c,b$\,, and the interference with the SM contribution has been neglected.
Here, the functions $w_q(s,{M_{Z^\prime}}^2)$ are independent of the $Z^\prime$ model and contain all information on parton distribution functions (PDFs) and QCD corrections. 
On the other hand, the coefficients
\begin{equation}
    c_q^f = 16 \pi^2  \alpha_4\,(F^2_{q_L}+F^2_{q_R})\, \mathcal{B}(Z^\prime \to f\bar{f})
\end{equation}
contain all model-dependent information, with $F_{q_{L,R}}$ being the $U(1)^\prime$ charges to the left- and right chiral components of the quark $q$.
Recently, the CMS collaboration also published limits on $M_{Z^\prime}$ as a function of $c_{u,d}^\ell$ where $\ell =e,\mu$ \cite{Sirunyan:2021khd}.
Adjusting $c_{u,d}^\ell$ to the BMs the $Z^\prime$ mass limits at 95 \% c.l. are 
\begin{equation}\label{eq:MassLimits}
    \begin{aligned}
    M_{Z^\prime}\big|_{\text{\BM 1}}  &\gtrsim 5.0 \text{ TeV}(5.4 \text{ TeV})\,,\\   
    M_{Z^\prime}\big|_{\text{\BM 2}}  &\gtrsim 5.9 \text{ TeV}(6.3 \text{ TeV})\,,\\
    M_{Z^\prime}\big|_{\text{\BM 3}}  &\gtrsim 5.8 \text{ TeV}(6.8 \text{ TeV})\,,
    \end{aligned}
\end{equation}
where the first and second value refer to scenarios with $2 M_{\psi}\ll M_{Z^\prime}$ and $2 M_{\psi} > M_{Z^\prime}$, respectively. 
Note however that in the analysis the limits from the dielectron and dimuon channel were combined whereas we do not induce decays to dielectrons. 
Hence, the actual mass limits should be a bit weaker than \eq{MassLimits}.
We stress that lepton flavor specific analyses are necessary  to further test the models discussed in this paper.
Moreover, searches for invisibly decaying high mass resonances, i.e. missing energy, offer a unique opportunity to search for the models, and deserve dedicated experimental analysis.
The sensitivity to models with $c_{u,d}^\ell \approx 0$ such as those with vanishing coupling to first and second generation quarks is much lower.
In particular, for \BM4 with  $(c_{u}^\ell,c_{d}^\ell) \sim (0, 10^{-7})$ no direct  mass bound such as  \eq{MassLimits} can be inferred  from the analysis \cite{Sirunyan:2021khd}. 
Note that for similar models without $Z'$ coupling to light quarks, bounds as low as $M_{Z'} > 1.2$~TeV were extracted \cite{Allanach:2019mfl,Allanach:2020kss,Allanach:2021gmj}. This serves us as a rough estimate while a more sophisticated recast of experimental data is beyond the scope of this work.

\begin{table*}[t]
\def\arraystretch{1.5}
\setlength\arrayrulewidth{1.3pt}
\centering
\begin{tabular}{|c|c|c|c|c|c|c|c|c|c|c|}\hline
Model & $\mu_0$ &$\alpha_4(\mu_0)$ & $\Cnine$ & $\Cten$  &  $Y^{u,d}_{ii}$ & $\ Y^{e}_{ii}\ $ & $\ Y^{\nu}_{ii}\ $ &    $r_{B_s}$ &  $\mathcal{B}(Z^\prime  \to \text{inv.})$ & $\ \nu_R\ $ 
\\
\hline
\BM 1 & 5~TeV & $1.87 \cdot 10^{-2}$  & $-$0.83 & 0   & $\checkmark$ & $\checkmark$ & $\mathsf{X}$  & 0.35  & 73\% & $\mathsf{X}$ 
\\
\BM 2 & 5~TeV & $5.97 \cdot 10^{-3}$ & $-$0.41 & $-\Cnine$  &  $\checkmark$ & $\mathsf{X}$ & $\mathsf{X}$ &  0.86 &  64\% & $\checkmark$ 
\\
\BM 3 & 5~TeV & $4.60 \cdot 10^{-2}$ & $-$0.71 & +0.24  &  $\checkmark$ & $\mathsf{X}$ & $\mathsf{X}$ &  0.60 &  87\% & $\checkmark$ 
\\ 
\BM 4 & 3~TeV & $2.46 \cdot 10^{-2}$ & $-$0.83 & 0  &  $\checkmark$ & $\checkmark$ & $\checkmark$ &  0.70 &  86\% & $\checkmark$ 
\\ 
\hline
\end{tabular}
\caption{Key features  of  Planck-safe benchmark models  (see \tab{Benchmarks} for charge assignments), showing
the values of    the matching scale, Wilson coefficients, the gauge coupling $\alpha_4$, 
 the  diagonal Yukawa interactions $(Y^{u, d,e,\nu}_{ii})$  which are generated (see \Sec{YukInv}), the branching ratio 
 for $Z'$ decays to missing energy, whether right-handed neutrinos $\nu_R$ are included, and
the room left by the $B_s$-mixing constraint $r_{B_s} \le 1$ before right-handed currents need to be invoked; $r_{B_s}= \left(|g_L^{bs}|^2/M^2_{Z^\prime}\right)/\left(|g_L^{bs}|^2/M^2_{Z^\prime}\right)_\text{max}$.
}
\label{tab:Overview_Benchmarks}	
\end{table*}

We stress that all our models can be probed at a future muon collider \cite{Delahaye:2019omf,Ali:2021xlw,Long:2020wfp,Zimmermann:2018wfu}. The $Z^\prime$ can be 
directly produced either on- or off-shell in the $s$-channel with a cross section enhanced by the large $U(1)^\prime$ coupling, and muon coupling $F_{L_2}$. In particular direct production with the dominant subsequent invisible decays offers a great discovery potential via $\mu^+ \mu^- \to Z^\prime \to \psi\bar{\psi},\nu \bar{\nu}$.
We obtain for invisible production to lowest order
\begin{equation}
    \begin{aligned}
    &\sigma (\mu^+ \mu^- \to Z^\prime \to \psi\bar{\psi},\nu \bar{\nu}) = \frac{s}{\left( M_{Z'}^2 - s\right)^2 +M_{Z^\prime}^2\Gamma_{Z^\prime}^2} \\
     & \ \cdot \frac{g_4^4 }{48 \pi }   \left(F_{L_2}^2 + F_{E_2}^2\right)  
     \left[  2 N_F F_\psi^2 + \sum_i (F_{L_i}^2 + F_{\nu_i}^2 ) \right],
    \end{aligned}
\end{equation}
which is about 880, 72, 560 and 4800  times larger in the respective models \BM{1-4} than the SM cross section
\begin{equation}
    \begin{aligned}
    &\sigma (\mu^+ \mu^- \to Z \to \nu \bar{\nu}) = \frac{3\, s \,\left(g_L^\nu\right)^2 \left[\left(g_L^\mu\right)^2 + \left(g_R^\mu\right)^2\right] }{48 \pi \left( M_{Z}^2 - s\right)^2}   
    \end{aligned}
\end{equation}
for $\sqrt{s}=3$ TeV \cite{Long:2020wfp}. 
Here the SM couplings to $Z$ are denoted as $g^\ell_{L,R} = (T_{3,\ell} - Q_\ell \sin^2 \theta_W)\cdot g_2/\cos \theta_W $.
Moreover, the interference between SM and BSM contribution as well as $t$-channel $W$-exchange for the  $\nu_\mu \bar{\nu}_\mu$ final state are neglected.  Note that in \BM4 for $\sqrt{s}=3$ TeV the $Z^\prime$ is produced resonantly. See \cite{Huang:2021nkl,Huang:2021biu} for recent muon collider studies of $Z'$ models explaining flavor anomalies.

On the other hand, bounds from neutrino trident production processes \cite{Altmannshofer:2014pba} are much weaker due to the large suppression $M_W^2/M_{Z'}^2$. In particular, the cross section \cite{Buras:2021btx} for all  our benchmark models acquire only a small deviations from the SM  $\sigma^\text{BM}/\sigma^\text{SM} \lesssim 10^{-2}$  which is within experimental uncertainties \cite{CHARM-II:1990dvf,CCFR:1991lpl,NuTeV:1998khj}.
Such constraints are typically more relevant for light $Z^\prime$ models aiming to explain the deviation of the muon anomalous magnetic moment $(g-2)_\mu$ \cite{Aoyama:2020ynm,Abi:2021gix}. In contrast, our models feature a heavy $Z^\prime$ with $M_{Z^\prime} \sim 5 \, \TeV \,(3 \TeV)$ and cannot account for $(g-2)_\mu$. In fact, after fixing $\alpha_4(M_{Z^\prime})$ in order to explain the $B$-anomalies, the $Z^\prime$ one-loop contribution to $(g-2)_\mu$ \cite{Leveille:1977rc,Lavoura:2003xp} is typically $2-3$ orders of magnitude too small. 
This might be interpreted as a hint that the $(g-2)_\mu$ discrepancy and the $B$-anomalies are actually generated by different types of new physics, see e.g. \cite{Greljo:2021npi,Wang:2021uqz} for a recent discussion.

\subsection{What if  \texorpdfstring{$C^\prime \neq 0$}{C' != 0}?   \label{sec:RH-fcnc}}

So far we considered $U(1)^\prime$ models inducing new physics Wilson coefficients $\Cnt$ with only left-handed quark couplings $g_L^{bs}$, which is sufficient to 
explain present $B$-data, cf. \tab{FitValues}. 
In these scenarios holds $R_K \simeq R_{K^\ast}$ \cite{Hiller:2014ula}. 
However, the latest experimental data from LHCb \cite{Aaij:2017vbb,Aaij:2021vac} suggest $R_K > R_{K^\ast}$ at a level of approximately $1\,\sigma$, see \tab{FitValues}.
If $R_K \neq R_{K^\ast}$ is confirmed with higher significance by future measurements it would indicate the presence of right-handed currents ($\Cninepr - \Ctenpr\neq 0$) as
\begin{equation}\label{eq:RKRatio}
   \frac{R_{K^\ast}}{R_{K}} \simeq 1 - 0.48 \, p \, (\Cninepr - \Ctenpr)
\end{equation}
at leading order in the BSM contribution, where the $K^\ast$ polarization fraction is $p \simeq 0.86$ \cite{Hiller:2014ula}. 
Hence, it is interesting to investigate the impact of $\Cntpr \neq 0$ on models.

In $Z^\prime$ extensions it generically holds 
\begin{equation}\label{eq:rDef}
    \frac{\Cninepr}{\Cnine} = \frac{\Ctenpr}{\Cten} = \frac{g_R^{bs}}{g_L^{bs}} \equiv r \,,
\end{equation}
see \eq{WCs}. 
Presently, the 4d fits favor $r<0$, see \tab{FitValues}, although they are also consistent with $r=0$. 

Assuming that $C_9^\mu <0$ remains a feature of the fit in the future and  that electrons remain SM-like, one finds that $r>0$
 (hence $C_9^{\prime \mu} <0$)  implies
$R_{K^*} > R_K$ whereas $r<0$ (hence $C_9^{\prime \mu} >0$) implies
$R_{K^*} < R_K$, and vice versa.
The cancellation of $Z^\prime$-contributions 
to $B_s$-mixing between left-handed and right-handed couplings, on the other hand, 
requires positive $r=X_s$, see \App{AppBSMixing}.
The $Z^\prime$ models can hence be probed, or ruled out, in the future using joint analysis of $b\to s \ell^+ \ell^-$ transition, muon-to-electron ratios
$R_{K^\ast}$ versus $R_K$, or similarly complementary ratios such as $R_\phi, R_{K_1}$, and friends~\cite{Hiller:2014ula}, and $B_s$-mixing.
Present data suggest that the window for the $U(1)^\prime$ models is tighter for $R_{K^\ast} < R_K$. 
Within the current $2\, \sigma$ theory uncertainty of $\Delta M_s^\text{SM}$ \cite{DiLuzio:2019jyq} we can roughly accommodate $R_{K^\ast} < R_K$ ($R_{K^\ast} > R_K$) at the level of $10\%\,(20\%)$ where the exact values depend on the benchmark scenario.

\section{\bf  Conclusions}
\label{sec:Conclusion}

We have put forward new flavorful $U(1)^\prime$ extensions of the SM that account for the $B$-anomalies
with an emphasis on  stability and predictivity up to  Planckian energies.
We have identified several benchmark models,
each  representing a viable global fit scenario to recent experimental data (\fig{c9c10_rk_benchmarks}),
and whose main features are summarized in Tab.~\ref{tab:Overview_Benchmarks}.
Further, our models can be extended naturally  
to accommodate  future measurements  (Sec~\ref{sec:RH-fcnc}).

While the $U(1)'$ interactions are central to explain the $B$-anomalies, we  found that new Yukawa, quartic, and portal interactions 
 are key to tame   Landau poles and to stabilize the  ground state including the notorious Higgs. 
Here, this is achieved with vector-like fermions, meson-like scalars, and right-handed neutrinos in some benchmarks (\tab{fields}). 
The demand for Planck safety has  also provided us with new  theory constraints 
 on low-energy  parameters (Figs.~\ref{fig:ParameterSpaceBM1},~\ref{fig:ParameterSpaceBM2},~\ref{fig:ParameterSpaceBM3},~\ref{fig:ParameterSpaceBM4}).
 In combination with constraints from phenomenology we observe
  an overall enhanced predictive power of models.

We have further studied the phenomenology of models, including an outlook on key collider signatures.
The most distinctive feature is that the $Z^\prime$ predominantly decays to invisibles, that is into new vector-like fermions, or, 
if this is kinematically closed, into neutrinos (\tab{Brs}).
Branching ratios in dineutrino modes including $B \to K^{(*)} \nu \bar \nu$ are only mildly enhanced in benchmarks $\lesssim 8\%$ relative to the predictions of the SM,  which is still within present theoretical uncertainties.
We look forward to further explorations of Planck-safe SM extensions, their phenomenology, and searches.
\\[4ex]

%\centerline{\bf Acknowledgments}
%\begin{acknowledgement}
\centerline{\bf Acknowledgments}
We are grateful to Hector Gisbert and Marcel Golz  for sharing results on the global $b \to s$ fits.
TH is supported by the \textit{Studienstiftung des Deutschen Volkes}.
DL is supported by the Science and Technology Facilities Council (STFC) under the Consolidated Grant ST/T00102X/1.
%\end{acknowledgement}

\setcounter{section}{0}
\section*{\bf Appendices}
\subsection{\texorpdfstring{$Z'$}{Z'} Models and Landau Poles}\label{sec:Landau}
In the introduction, we have argued that successful explanations of the $B$-anomalies with $U(1)'$ models 
 generically come bundled with the loss of predictivity  below the Planck scale. 
Here, we detail the derivation of the bound \eq{Landau}.
We recall that explaining $R_{K^{(*)}}$ with  a tree-level mediator and couplings  of order unity points towards a  scale of about $40$ TeV \cite{Hiller:2021pul}.
Moreover, a generic lower bound from LHC searches on a heavy (electroweak mass or above) $Z^\prime$ mass is roughly $5$~TeV \cite{Sirunyan:2021khd}.
Minimal  $U(1)^\prime$ charge assignments which can explain the $B$-anomalies
are to the left-handed muon $(F_{L_2})$ and to the left-handed $b$-quark $(F_{Q_3})$, or,
alternatively, to  the left-handed muon and to the left-handed $s$-quark.
The FCNC quark vertex also requires flavor mixing given by
a CKM factor $V_{tb} V_{ts}^*$ when demanding the CKM rotation to be in the down-sector, i.e. $V_d \simeq V_\text{CKM}$ and $V_u\simeq \mathds{1}$. Altogether, and in terms of $\alpha_4=g_4^2/(4 \pi)^2$,  
 we obtain the estimate
\begin{equation}\label{eq:alpha4bound}
\begin{aligned}
\alpha_4 (\mu_0=5 \text{TeV}) &\sim   
\frac{1}{(4\pi)^2}\frac{(5 \text{TeV}/40 \text{TeV})^2}{V_{tb}^{\phantom{*}} V_{ts}^* F_{L_2} F_{Q_3} }\\
&\sim \frac{1}{40\pi^2}\frac{1}{F_{L_2} F_{Q_3}}\,.
\end{aligned}
\end{equation}
A smaller $\alpha_4$
can be achieved with larger $bs$-mixing angles than in \eq{Q2313}. 
This, however, would require large cancellations between up- and down-quark flavor rotations (see \App{chargeMixDetails}). 

On the other hand, the leading-order renormalization group running  for $\alpha_4 (\mu)$ is given by 
\begin{equation}\label{eq:flow}
\frac{d \alpha_4}{d \ln \mu } =  B_4 \,\alpha_4^2\,,
\end{equation}
where subleading  effects from gauge-kinetic mixing have been neglected. 
Integrating \eq{flow} identifies the Landau pole from $1/\alpha_4(\mu_L)=0$, hence
\begin{equation}
    \ln \mu_L/\mu_0 = \frac{1}{B_4 \alpha_4(\mu_0)}\,.
\end{equation}
With the left-handed muon and  $b$-quark transforming under the $U(1)'$, and a minimal amount of  
extra charges to avoid gauge anomalies, we find the bound
\begin{equation}\label{eq:B4-min}
    B_4 \gtrsim \tfrac{16}3(F_{L_2}^2 + 3 F_{Q_3}^2)\,.
\end{equation}
Together with \eq{alpha4bound}, we have
\begin{equation}
\frac{1}{B_4 \alpha_4(\mu_0)}= \frac{15\pi^2}{2} \frac{F_{L_2}F_{Q_3}}{F_{L_2}^2+3F_{Q_3}^2} \lesssim\frac{15\pi^2}{4\sqrt{3}}\,,
 \end{equation}
where the inequality holds true irrespective of the charge assignments $F_{L_2}$ and $F_{Q_3}$.
Quantitatively, this implies an upper bound for the scale of the Landau pole
\begin{equation}\label{eq:AppLandau}
\mu_{\rm L}\lesssim  10^{10}\, {\rm TeV}\ll M_{\rm Pl}
\end{equation}
stated in \eq{Landau}.  
As soon as beyond-minimal $U(1)'$ charge carriers are present, the  Landau pole is shifted towards lower energies,  often significantly ($e.g.$~our benchmark models with or without $\psi$, $S$ and $\phi$ fields).

It is worth noting that the quantum consistency of models (anomaly cancellation) has been  important to achieve the bounds. 
We have also checked that gauge-kinetic mixing, which is severely constrained at the matching scale  from electroweak precision  data, \eq{eta_bound}, has a negligible impact on the  result  $\mu_{\rm L}\ll M_{\rm Pl}$.  

Let us briefly comment on how 
the generic bound  \eq{AppLandau} can be lifted in fine-tuned settings.
Firstly, larger $bs$-mixing angles can lower  \eq{alpha4bound} and enhance the scale $\mu_L$. 
This requires   large cancellations between $V_d$ and $V_u$ (we have discarded this possibility for our study).
A second option relates to lower values of  $M_{Z^\prime}$. 
These become available in models with   first and second generation quarks being  uncharged under the $U(1)^\prime$, such
that $pp$-production is strongly suppressed and sensitivity in LHC searches   lost \cite{Sirunyan:2021khd,ATLAS:2019erb}.
This scenario has been explored in \BM4, which further constitutes a minimal model for explaining the $B$-anomalies with $V_d \simeq V_\text{CKM}$ and $V_u \simeq \mathds{1}$. However, even in minimal settings (e.g.~after decoupling the fields $\psi,\, S,\, \phi$)  we still find  a sub-planckian Landau pole   for  $M_{Z^\prime} \gtrsim 1.8$~TeV. 
Finally, explaining $B$-anomalies with  substantially lighter $Z'$s may still be possible in  fine-tuned models, see e.g.~\cite{Allanach:2020kss}.
In these cases the Landau pole could be trans-planckian, and one is left to deal with the in- or  metastability  of the quantum vacuum including the Higgs.

\subsection{Flavor Rotations in the Quark Sector }\label{sec:chargeMixDetails}

The FCNC couplings $g_{L,R}^{bs}$ are generated by rotations from gauge to mass basis. In the quark-sector, four unitary rotations exist, those for up(down)-singlets $U_u(U_d)$ and up(down)-doublets $V_u(V_d)$, all of which are apriori (without a theory of flavor) unconstrained except for  the product $V_u^\dagger\,V_d=V_\text{CKM}$,
where $V_\text{CKM}$ is the CKM matrix.
In this work we consider rotations in the down-sector, $V_u\approx \mathds1\,,U_u\approx \mathds1$, which yields in particular $V_d\approx V_\text{CKM}$, to maximize effects in the down sector.
 Note also that we discard the possibility of large cancellations between up- and down-quark flavor rotations,
 corresponding to large mixing angles.
 
To study mixing in the left-handed sector we write the flavor structure of  $U(1)^\prime$ charges  in the gauge basis
\begin{align}
	\mathcal{Q} = \begin{pmatrix}
	F_{Q_1} & 0 & 0 \\
	0 & F_{Q_2} & 0 \\
	0 & 0 & F_{Q_3} \\
	\end{pmatrix} \,.
\end{align}
With CKM-mixing residing in the down-sector, we can rotate to the down mass basis via
\begin{align}
	\mathcal{Q}^\prime = V_\text{CKM}^\dagger\,\mathcal{Q}\,V_\text{CKM} \,.
\end{align}
One obtains for  the $bs$ and $bd$ vertices
\begin{align}\label{eq:Q2313}
\begin{split}
      \mathcal{Q}^\prime_{23} &= V_{tb}\,V_{ts}^\ast\left(F_{Q_3}-F_{Q_1}\right) +V_{cb}\,V_{cs}^\ast\left(F_{Q_2}-F_{Q_1}\right) \\
      &\approx V_{tb}\,V_{ts}^\ast\left(F_{Q_3}-F_{Q_2}\right)\,, \\  \\
      \mathcal{Q}^\prime_{13} &= V_{tb}\,V_{td}^\ast\left(F_{Q_3}-F_{Q_1}\right) +V_{cb}\,V_{cd}^\ast\left(F_{Q_2}-F_{Q_1}\right) \\
      &\approx V_{tb}\,V_{td}^\ast\left(F_{Q_3}-F_{Q_2}\right)\,,     
\end{split}
\end{align}
respectively,
where CKM unitarity and hierarchy has been used in each of the last steps.
It follows that
\begin{align}\label{eq:gbsbd}
g^{bs}_L&= \mathcal{Q}^\prime_{23}\,g_4 \,, \\
g^{bd}_L&= \mathcal{Q}^\prime_{13}\,g_4 \,.
\end{align}
For the diagonal couplings we find $g_L^{ii}\approx F_{Q_i}\,g_4$, where $i$ denotes the generation index.

For the right-handed quarks we define
\begin{align}
	\mathcal{D} = \begin{pmatrix}
	F_{D_1} & 0 & 0 \\
	0 & F_{D_2} & 0 \\
	0 & 0 & F_{D_3} \\
	\end{pmatrix} \,,
\end{align}
in the gauge basis, where we again assume mixing only in the down-sector.
We rotate to the mass basis via
\begin{align}
	\mathcal{D}^\prime = U_d^\dagger\,\,\mathcal{D}\, U_d \,,
\end{align}
where $U_d$ describes the unitary transformation.
We parametrize $U_d$ as
\begin{align}
U_d = \begin{pmatrix}
1 & 0 & 0 \\
0 & \cos{\theta_d} & -e^{\text{i} \phi_R}\,\sin{\theta_d} \\
0 & e^{-\text{i} \phi_R}\,\sin{\theta_d} & \cos{\theta_d}  \\
\end{pmatrix} 
\end{align}
where $\theta_d$ is $bs$-mixing angle for the down-type quark singlets and $\phi_R$ the corresponding CP-phase.
Here, we have neglected off-diagonal mixing of  the first generation quark singlets, such that, e.g., $g_R^{bd}=0$. We obtain
\begin{align}
    g_R^{bs} &=\mathcal{D}^\prime_{23}\,g_4 = \left(F_{D_3} -F_{D_2} \right)\sin{\theta_d}\cos{\theta_d}\,e^{\text{i}\phi_R}\,g_4 \,,\\
    g_R^{bd}&=\mathcal{D}^\prime_{13}\,g_4= 0 \,.
\end{align}
Since in this work we are not interested in CP-violation, we set $\phi_R=0$ and, hence,  obtain real-valued  $g_R^{bs}$.

The generation-diagonal couplings read
\begin{align}
  g_R^{bb} &=\mathcal{D}^\prime_{33}\,g_4 = \left(F_{D_3}\cos^2{\theta_d} + F_{D_2}\sin^2{\theta_d}\right)\,g_4 \,,\\
  g_R^{ss} &=\mathcal{D}^\prime_{22}\,g_4 = \left(F_{D_2}\cos^2{\theta_d} + F_{D_3}\sin^2{\theta_d}\right)\,g_4 \,.
\end{align}
For $\theta_d \ll 1$ contributions from other-generation charges can be neglected, that is, $g_R^{ii} \approx F_{D_i}\,g_4$.

To suppress $B_s$-mixing contributions  from the $Z^\prime$ (see \App{AppBSMixing} for details) one could fix
\begin{align}
\begin{split}
        \sin2\theta_d &= \frac{2 V_{tb} V_{ts}^*}{X_s} \frac{F_{Q_3}-F_{Q_2}}{F_{D_3}-F_{D_2}} \\
        &\approx ( -8 \times 10^{-3}) \cdot \frac{F_{Q_3}-F_{Q_2}}{F_{D_3}-F_{D_2}} 
\end{split}
\end{align}
where in the last step we assumed $X_s\approx 10$. Hence, $\theta_d$ is small in models with $F_{Q_3}-F_{Q_2}$ of the same order or smaller as $F_{D_3}-F_{D_2}$.

\subsection{Evading \texorpdfstring{$B$}{B}-mixing Constraints}
\label{sec:AppBSMixing}

We describe the effects of $B_q$-mixing, $q={d,s}$ via the effective Hamiltonian at low energies \cite{DiLuzio:2019jyq}:
\begin{align}
\begin{split}
      \mathcal{H}_\text{eff}^{\Delta B=2} \supset &\frac{4\,G_F}{\sqrt{2}}\,\left(V_{tb}\,V_{tq}^\ast\right)^2\bigg[  C_{bq}^{LL}\,\left( \bar q_L \gamma^\mu b_L \right)^2 \\
     &+ C_{bq}^{RR}\,\left( \bar q_R \gamma^\mu b_R \right)^2 \\
    &+ C^{LR}_{bq}\left( \bar q_L \gamma^\mu b_L \right)\left(\bar q_R \gamma^\mu b_R \right) \bigg] +\text{h.c.} \,,  
\end{split}
\end{align}
with 
\begin{align}
\begin{split}
    C_{bq}^{LL} &= \frac{1}{4\sqrt{2}\,G_F}\frac{1}{\left(V_{tb}\,V_{tq}^\ast \right)^2}\,\left(g_L^{bq}\right)^2 \,, \\
    C_{bq}^{RR} &= \frac{1}{4\sqrt{2}\,G_F}\frac{1}{\left(V_{tb}\,V_{tq}^\ast \right)^2}\,\left(g_R^{bq}\right)^2 \,, \\
    C_{bq}^{LR} &= \frac{\sqrt{2}}{4\,G_F}\frac{1}{\left(V_{tb}\,V_{tq}^\ast \right)^2}\,g_L^{bq}\,g_R^{bq} \,. 
\end{split}
\end{align}
The SM predictions as well as the experimental values of the mass differences of the $B$ mesons \cite{DiLuzio:2019jyq}
are taken within their $2.5\,\sigma$ uncertainties to obtain the total (SM+NP) contribution normalized to the SM as 
\begin{align}\label{eq:dMsd+NP}
\frac{\Delta M_{s(d)}^\text{SM+NP}}{\Delta M_{s(d)}^\text{SM}} \leq 1.156\,(1.154)\,,
\end{align}
where NP effects can be as large as $15\,\%$ compared to the SM.
To study contributions from $B_s$-mixing via a $Z^\prime$ boson we write
\begin{align}
    &\frac{\Delta M_s^\text{SM+NP}}{\Delta M_s^\text{SM}} = \bigg| 1 + \frac{A^{6/23}}{R_\text{loop}^\text{SM}}\bigg\lbrace C_{bs}^{LL} + C_{bs}^{RR} \notag \\
    &- \frac{1}{2\,A^{3/23}}\bigg\lbrack \frac{B_5}{B_1}\,\left( \frac{m_{B_s}^2}{(m_b+m_s)^2}+\frac{3}{2}\right) \\
    &+\frac{B_4}{B_1}\,\left( \frac{m_{B_s}^2}{(m_b+m_s)^2}+\frac{1}{6}\right)\left( A^{-27/23}-1\right) \bigg\rbrack\,C_{bs}^{LR}
    \bigg\rbrace\bigg| \,, \notag
\end{align}
where running effects from the NP scale $\mu_\text{NP} \approx m_{Z^\prime}$ to $\mu_b$ of the coefficients $C_{bs}$ are taken into account and generate the contributions $\sim C^{LR}_{bs}$ with bag parameters $B_{4,5}$ appearing in this expression. 
Here, $A=\alphas(\mu_\text{NP})/\alphas(\mu_b)$ 
and the loop function
\begin{align}
    R_\text{loop}^\text{SM} = \frac{\sqrt{2}G_F\,M_W^2\,\hat{\eta}_B\,S_0(x_t)}{16\,\pi^2} \,,
\end{align}
with $\hat\eta_B \approx 0.84$, and bag parameters $B_i$ defined in Ref.~\cite{DiLuzio:2019jyq}, where also 
weighted averages of these parameters for $B_{s,d}$-mixing are provided.

It follows that
\begin{align} \label{eq:ZprimeDeltaMs}
\begin{split}
    \frac{\Delta M_s^\text{SM+NP}}{\Delta M_s^\text{SM}} = &\bigg| 1 + 200\left(\frac{5\,\TeV}{M_{Z^\prime}}\right)^2\\
    &\cdot \bigg\lbrack \left(g_L^{bs}\right)^2 + \left(g_R^{bs}\right)^2 - X_s \,g_L^{bs}\,g_R^{bs}
    \bigg\rbrack\bigg| \,,
\end{split}    
\end{align}
with $X_s\approx 10$ and assuming $m_{Z^\prime} \approx 5\,\TeV$.
We find that for 
\begin{align}
    \left(g_L^{bs}\right)^2 + \left(g_R^{bs}\right)^2 -X_s\,g_L^{bs}\,g_R^{bs} = w\,, 
\end{align}
with $w \approx 0$ (very small deviations allowed from mixing constraints) the effect of NP in $B_s$-mixing is minimized.
This can be translated into 
(in analogy to Ref.~\cite{Bause:2019vpr},\,Eq.~(B10) for charm)
\begin{equation}
g_L^{bs} = g_R^{bs}\left( \frac{X_s}{2} \pm \sqrt{\left(\frac{X_s^2}{4}-1\right) + \frac{w}{\left(g_R^{bs}\right)^2}}\right)
\end{equation}
which is symmetric in $L\leftrightarrow R$ and implies
\begin{align}
&g_L^{bs} = g_R^{bs}\,X_s \,\,(\text{LH-dominated}) \\
\lor \quad &g_L^{bs} = g_R^{bs}\,\frac{1}{X_s}\,\,(\text{RH-dominated}) \,,
\end{align}
where we employed $4/X_s^2 \ll 1$ and $w/(g_R^{bs})^2 \ll 1$ in the last step. 
Note that the LH-dominated scenario is presently preferred by $R_{K^{(\ast)}}$ data, see \tab{FitValues}.
Imposing this hierarchy we can circumvent the $B_s$-mixing bound. 
We thereby induce semileptonic Wilson coefficients 
\begin{align}
    \frac{C_{9,10}^\ell}{C^{\prime \ell}_{9,10}} &= X_s \quad (\text{LH-dominated})\,, \label{eq:LHcancel} \\
    \frac{C_{9,10}^\ell}{C^{\prime \ell}_{9,10}} &= \frac{1}{X_s} \quad (\text{RH-dominated})\,,     
\end{align}
with the same hierarchy as between $g_L^{bs}$ and $g_R^{bs}$.

Constraints from  $\Delta M_d$ and $\Delta M_s$ on our models are similar  due to similar room in both sectors \eq{dMsd+NP}
and the fact that the  $Z^\prime$ induced contributions $(g_L^{bq})^2\sim (V_{tb}  V_{tq}^*)^2$ follow the CKM-suppression of the SM, cf.~\App{chargeMixDetails}.

%\vspace{0.4cm}
\subsection{When and Why: Right-Handed Neutrinos}
\label{sec:AppRHN}
If no $U(1)'$ charged right-handed neutrinos were present ($F_{\nu_i}= 0$), solving the anomaly cancellation conditions \eq{ACCs}, while also imposing gauge invariance of diagonal quark Yukawas \eq{QuarkMassYCCs}, as well as constraints from Kaon mixing \eq{KaonBound} and electron couplings \eq{ElectronBound} yields
\begin{equation}\label{eq:chargeConNoRHneutrinos}
\begin{aligned}
    &  F_H = 0\,,   \quad  F_{q_3} = -2\,F_{q_1} = -2\,F_{q_2}\,,\\
    & F_{L_3} = -F_{L_2}\,, \quad F_{E_3} = - F_{E_2} = \mp F_{L_2}\,,  
\end{aligned}
\end{equation}
where $q_i = Q_i$, $U_i$, $D_i$. In order to generate the  hierarchy of Wilson coefficients \eq{C9-C10-pattern} in accord with \eq{BAnomaliesChargeConstr}, only the case $F_{E_2} = F_{L_2}$ in \eq{chargeConNoRHneutrinos} is viable.
While this allows for NP scenarios with $C^\mu_{9} \neq 0$,  
it also stipulates $C^\mu_{10} \sim (F_{E_2} - F_{L_2} )= 0$, cf.~\eq{WCs} and \eq{gll}.
Therefore, $C^\mu_{10} \neq 0$ requires  right-handed neutrinos, or more generally,  fermions that are chiral and charged under the SM gauge group and/or $U(1)'$ that enter ACCs \eq{ACCs} and relax \eq{chargeConNoRHneutrinos}. 
Hence, right-handed neutrinos are mandatory for \BM{2,3}, where $C^\mu_{10} \neq 0$. 
Moreover, right-handed neutrinos are also required for the minimal \BM4 in order to couple $Z'$ to the third quark generation only, as implied by \eq{chargeConNoRHneutrinos}. On the other hand, \BM1 is both compliant with $C^\mu_{10} = 0$ as well as with~\eq{chargeConNoRHneutrinos}, eliminating the need for right-handed neutrinos.

Let us briefly comment on what happens in more minimal models where \eq{QuarkMassYCCs} is only enforced for the third generation quarks.
Here the absence of right-handed neutrinos is less restrictive. For instance, it does not result in a strict requirement for $F_H$, as opposed to \eq{chargeConNoRHneutrinos} which predicts \eq{FH=0}. 
Moreover, new physics scenarios with either $C^\mu_{10} = 0$ or $C^\mu_{10} \neq 0$ are accessible.

%%%%%%%%%%%%%%%%%%%%%%%%%%%%%%%
\bibliographystyle{JHEP}
\bibliography{ref-bib.bib}

\end{document}